\documentclass[prd,nofootinbib,preprint,superscriptaddress,twocolumn,10pt]{revtex4}
\usepackage{amsmath,amssymb}
\usepackage{soul}
\usepackage{epsfig}
\usepackage{graphicx}
\usepackage[usenames,dvipsnames]{color}
\usepackage{slashed}
\usepackage[colorlinks,citecolor=blue]{hyperref}
\usepackage{color}
\begin{document}
\title{Asymmetric Self-interacting Dark Matter via Dirac Leptogenesis}

%\author{abc}
\author{Manoranjan Dutta}
\email{ph18resch11007@iith.ac.in}
\affiliation{Department of Physics, Indian Institute of Technology
Hyderabad}
\author{Nimmala Narendra}
\email{nnarendra@prl.res.in}
\affiliation{Theoretical Physics Division, Physical Research Laboratory, Ahmedabad - 380009, India}
\author{Narendra Sahu}
\email{nsahu@phy.iith.ac.in}
\affiliation{Department of Physics, Indian Institute of Technology
Hyderabad}
\author{Sujay Shil}
\email{sujay.s@iitgn.ac.in}
\affiliation{IIT Gandhinagar, Palaj Campus, Gujarat 382355, India}
\affiliation{Institute of Physics, Sachivalaya Marg, Bhubaneswar, Pin-751005, Odisha}
\affiliation{Homi Bhabha National Institute, BARC Training School Complex, Anushakti Nagar, Mumbai 400094, India}

%%%%%%%%%%%%%%%%%%%%%%%%%%%%%%%%

%\input mylatex

%\doublespacing
%\textwidth 15.55cm \textheight 22.5cm
%\hoffset -2cm
%\voffset -1cm

%\def\jw#1{{\color{blue}{#1}}}
%\def\jwn#1{{\color{red}{\bf[#1]}}}

%================================================
%==================================================

%\begin{document}
%\maketitle
%\newpage
%\title{Asymmetric Self-interacting Dark Matter via Dirac Leptogenesis}

%\author{abc}
%\author[a]{Manoranjan Dutta}
%\emailAdd{ph18resch11007@iith.ac.in}

%\author[b]{Nimmala Narendra}
%\emailAdd{nnarendra@prl.res.in}

%\author[a]{Narendra Sahu}
%\emailAdd{nsahu@phy.iith.ac.in}

%\author[c,d,c]{Sujay Shil}
%\emailAdd{sujay.s@iitgn.ac.in}

%\affiliation[a]{Department of Physics, Indian Institute of Technology Hyderabad}

%\affiliation[b]{Theoretical Physics Division, Physical Research Laboratory, Ahmedabad - 380009, India}

%\affiliation[c]{Department of Physics, Indian Institute of Technology Hyderabad}

%\affiliation[c]{IIT Gandhinagar, Palaj Campus, Gujarat 382355, India}
%\affiliation[d]{Institute of Physics, Sachivalaya Marg, Bhubaneswar, Pin-751005, Odisha}
%\affiliation[e]{Homi Bhabha National Institute, BARC Training School Complex, Anushakti Nagar, Mumbai 400094, India}

%\keywords{}
%%%%%%%%%%%%%%%%%%%%

%\flushbottom

%%%%%%%%%%%%%%%%%%%%%%%%%%%%%%%%%%%%%%%%%%%%%%%%%%%%%%%%%%%%%%%%%%%%%
  
%%%%%%%%%%%%%%%%%%%%%%%%%%%%%%%%%%%%%%%%%%%%%%%%%%%%%%%%%%%%%%%%%%%%%%%  
\begin{abstract}
The nature of neutrinos, whether Dirac or Majorana, is hitherto not known. Assuming that the neutrinos are Dirac, which needs $B-L$ 
to be an exact symmetry, we make an attempt to explain the observed proportionality between the relic densities of dark matter (DM) 
and baryonic matter in the present Universe ${\it i.e.,}\,\, \Omega_{\rm DM} \approx 5\, \Omega_{\rm B}$. We extend 
the Standard Model (SM) by introducing heavy scalar doublets $X_i, i=1,2$ and $\eta$, two singlet scalars $\Phi$ and $\Phi'$, 
a vector-like Dirac fermion $\chi$ representing the DM and three right-handed neutrinos $\nu_{R_i}, i=1,2,3$. Assuming $B-L$ is an exact 
symmetry of the early Universe, the CP-violating out-of-equilibrium decay of heavy scalar doublets: $X_i, i=1,2$ to the SM lepton 
doublet $L$ and the right-handed neutrino $\nu_R$, generate equal and opposite $B-L$ asymmetry among left ($\nu_L$) and right ($\nu_R$)-handed neutrinos.
%Assuming the existence of heavy $SU(2)_L$ scalar doublet $(X= (X^0, X^-)^T)$ 
%in the early Universe, an equal and opposite $B-L$ asymmetry can be generated in left and right-handed sectors  by the CP-violating out-of-equilibrium 
%decay $X^0 \to \nu_L \nu_R$ since $B-L$ is an exact symmetry. 
We ensure that $\nu_L-\nu_R$ equilibration does not occur until below the electroweak (EW) 
phase transition during which a part of the lepton asymmetry gets converted to dark matter asymmetry through a dimension eight operator, which conserves $B-L$ symmetry and remains in thermal equilibrium above sphaleron decoupling temperature. A part of the remaining $B-L$ asymmetry then gets converted to a net B-asymmetry through EW-sphalerons which are active at a temperature above 100 GeV. To alleviate the small-scale anomalies of $\Lambda$CDM, we assume 
the DM ($\chi$) to be self-interacting via a light mediator $\Phi$, which not only depletes the symmetric component of the DM, but also paves a way to detect the DM at terrestrial laboratories through $\Phi-H$ mixing, where $H$ is the SM Higgs doublet.
\end{abstract}

\maketitle
\newpage

%%%%%%%%%%%%%%%%%%%%%%%%%%%%%%%%%%%%%%%%%%%%%%%%%%%%%%%%%%%%%%%%%%%%%%%
%\preprint{ HRI-RECAPP-2020-007}
%\keywords{Dark matter, asymmetric dark matter, self-interacting dark matter, leptogenesis, neutrino mass.}

%\begin{document}

%\maketitle
%\flushbottom

\section{Introduction} \label{Intro}
%%%%%%%%%%%%%%%%%%%%%%%%%%%%%%%%%%%%%%%%%%%%5
It is presumed that the early Universe has gone through a period of exponential expansion called inflation to solve 
the cosmological problems. At the end of inflation, the Universe is reheated to give rise to a thermal bath with reheating 
temperature $T_R \gtrsim 4 $MeV in order to facilitate the big-bang nucleosynthesis (BBN). It is expected that the different 
components of the present Universe, such as dark matter, dark energy and baryonic matter must have been cooked in a 
post-inflationary thermal bath. At present, the visible component (baryonic matter) of the Universe is best understood 
in terms of the standard model (SM) of particle physics, which is based on the gauge group $SU(3)_c \times SU(2)_L 
\times U(1)_Y$. However, the SM fails to explain many other aspects of the observed Universe, such as baryon asymmetry, 
dark matter, dark energy, non-zero but small neutrino mass etc.  

Within the framework of the SM, neutrinos are exactly massless. However, the solar and atmospheric neutrino oscillation experiments 
hint towards the non-zero masses and mixings of light neutrinos. In fact, this has been further confirmed by relatively 
recent oscillation experiments like T2K~\cite{Abe:2011sj, Abe:2013hdq}, Double Chooz~\cite{Abe:2011fz, DoubleChooz:2019qbj}, 
Daya Bay~\cite{An:2012eh, An:2017osx, Adey:2018zwh}, Reno~\cite{Ahn:2012nd} and MINOS~\cite{Adamson:2013whj, Adamson:2014vgd}. 
For a recent global fit of neutrino oscillation experiment, we refer to \cite{global_fit_data}. The neutrino 
masses are also further constrained by cosmology. The cosmic microwave background radiation data give an upper bound on the sum of light 
neutrino masses to be $\sum_i|m_i| < 0.12$ eV \cite{Vagnozzi:2017ovm}. Thus the data from various sources imply that neutrinos 
have mass. However, the nature of neutrino mass, whether Dirac or Majorana, is not confirmed yet. If the neutrinos are Majorana, which 
implies lepton number is violated by two units, then the seesaw mechanisms: type-I~\cite{Minkowski:1977sc, Yanagida:1981xy, GellMann:1980vs, 
Mohapatra:1979ia}, type-II~\cite{type_II_seesaw_1, type_II_seesaw_2, type_II_seesaw_3, type_II_seesaw_4, type_II_seesaw_5,type_II_seesaw_6,
type_II_seesaw_7}, type-III~\cite{type_III_seesaw} or their variants~\cite{ray_volkas_review} are the best theoretical candidates to get 
their sub-eV masses. It is important to note that all these seesaw mechanisms and their variants introduce additional heavy particles 
to the SM. After integrating out the heavy degrees of freedom we get the effective dimension five operator $LLHH/\Lambda$, where $L, H$ 
are lepton and Higgs doublets of the SM and $\Lambda$ is the mass scale of heavy degrees of freedom introduced in the various seesaw 
mechanisms. After electroweak phase transition, the dimension five operator generates sub-eV Majorana masses of neutrinos. On the other 
hand, if the neutrinos are Dirac, then lepton number is an exact symmetry of nature as it stands now in the SM. In this case, the sub-eV 
masses of light neutrinos imply that the Yukawa coupling involving $\bar{L}\widetilde{H} N_R$, with $N_R$ being the singlet right-handed 
neutrino, is of $\mathcal{O}(10^{-12})$, which is almost six orders of magnitude less than the electron Yukawa coupling. Thus the sub-eV 
Dirac mass of light neutrinos requires substantial fine-tuning.     

Another important aspect of the SM is the identity of dark matter, which plays a major role throughout the evolution of 
the Universe. Astrophysical evidences from galaxy rotation curve, gravitational lensing and large scale structure of the 
Universe confirmed the existence of dark matter\cite{dm_review_1,dm_review_2}. In fact, the satellite-borne experiments 
WMAP\cite{Hinshaw:2012aka} and Planck\cite{Akrami:2018vks}, which measure the temperature fluctuation in the cosmic 
microwave background (CMBR), precisely determine the relic abundance of baryon and dark matter (DM) to be $\Omega_{\rm B} h^{2}=0.02237 
\pm 0.00015$ and $\Omega_{\rm DM} h^{2}=0.1200 \pm 0.0012$ respectively at 68\% CL, where $\Omega_{\rm DM}$ is the density 
parameter and $h = \text{Hubble Parameter}/(100 \;\text{km} ~\text{s}^{-1} \text{Mpc}^{-1})$ is the reduced Hubble constant. This implies 
that the DM abundance is about five times the baryon abundance: {\it i.e.} $\Omega_{\rm DM} \approx 5 \Omega_{\rm B}$. While it is very 
natural for the Universe to start in a baryon symmetric manner, the present Universe is highly baryon asymmetric, giving rise to the 
long-standing puzzle of the baryon asymmetry of the Universe (BAU).  The observed BAU is quantitatively expressed by the ratio of baryon 
density over anti-baryons density to photon density as \cite{Aghanim:2018eyx}, 
\begin{equation}
\eta_B = \frac{n_{B}-n_{\overline{B}}}{n_{\gamma}} \simeq 6.2 \times 10^{-10}.
\label{etaBobs}
\end{equation}
%The quoted value of baryon to photon ratio based on the cosmic microwave background (CMB) measurements, agrees with the big bang nucleosynthesis (BBN) estimates as well ~\cite{Zyla:2020zbs}. 
The origin of this asymmetry is also not known along with the particle nature of DM.

In the well-established $\Lambda{\rm CDM}$ model, DM is assumed to be cold and collisionless which is supposed to have facilitated the structure 
formation in the early Universe by providing the necessary gravitational potential for primordial density fluctuations to grow. However, cosmological 
simulations in recent times reveal a few severe discrepancies of the $\Lambda{\rm CDM}$ model at small scales, leading to anomalies such as 
the cusp-core problem, missing satellite problem and too-big-to-fail problem \cite{Tulin:2017ara, Bullock:2017xww}. To alleviate these 
small-scale $\Lambda{\rm CDM}$ anomalies, Spergel and Steinhardt proposed in 2000 an interesting alternative to cold dark matter (CDM) in 
terms of self-interacting dark matter (SIDM) \cite{Spergel:1999mh}. Earlier studies in this direction can be found in \cite{Carlson:1992fn, deLaix:1995vi}. 
SIDM can have large self-scattering cross-sections of $\mathcal{O} (10^{-24} {\rm cm^2/GeV})$ \cite{Buckley:2009in, Feng:2009hw, Feng:2009mn, Loeb:2010gj, Zavala:2012us, Vogelsberger:2012ku}, which is way too larger than typical cross-section of $\mathcal{O} (10^{-38} {\rm cm^2/GeV})$ for weakly-interacting 
massive particles (WIMPs), a well-suited class of candidates for CDM scenario. Such large self-interacting cross-sections of DM can be naturally 
realized in scenarios where DM has a light mediator. In such a scenario, self-interaction is stronger for smaller DM velocities such that it can 
have a large impact on small scale structures, while it gets reduced at larger scales due to large velocities of DM and hence remains consistent 
with large scale CDM predictions \cite{Buckley:2009in, Feng:2009hw, Feng:2009mn, Loeb:2010gj, Bringmann:2016din, Kaplinghat:2015aga, Aarssen:2012fx, 
Tulin:2013teo}. The light mediator also mixes with the SM Higgs paving a way for detecting DM at direct detection experiments \cite{Kaplinghat:2013yxa, DelNobile:2015uua}. Several model building efforts have been made to realize such scenarios, see \cite{Kouvaris:2014uoa, Bernal:2015ova, Kainulainen:2015sva, Hambye:2019tjt, Cirelli:2016rnw, Kahlhoefer:2017umn, Dutta:2021wbn, Borah:2021yek,Borah:2021pet,Borah:2021rbx,Borah:2021qmi} and references therein.
\begin{figure}[ht]
				\centering
				\includegraphics[height = 5.5cm,width=8cm]{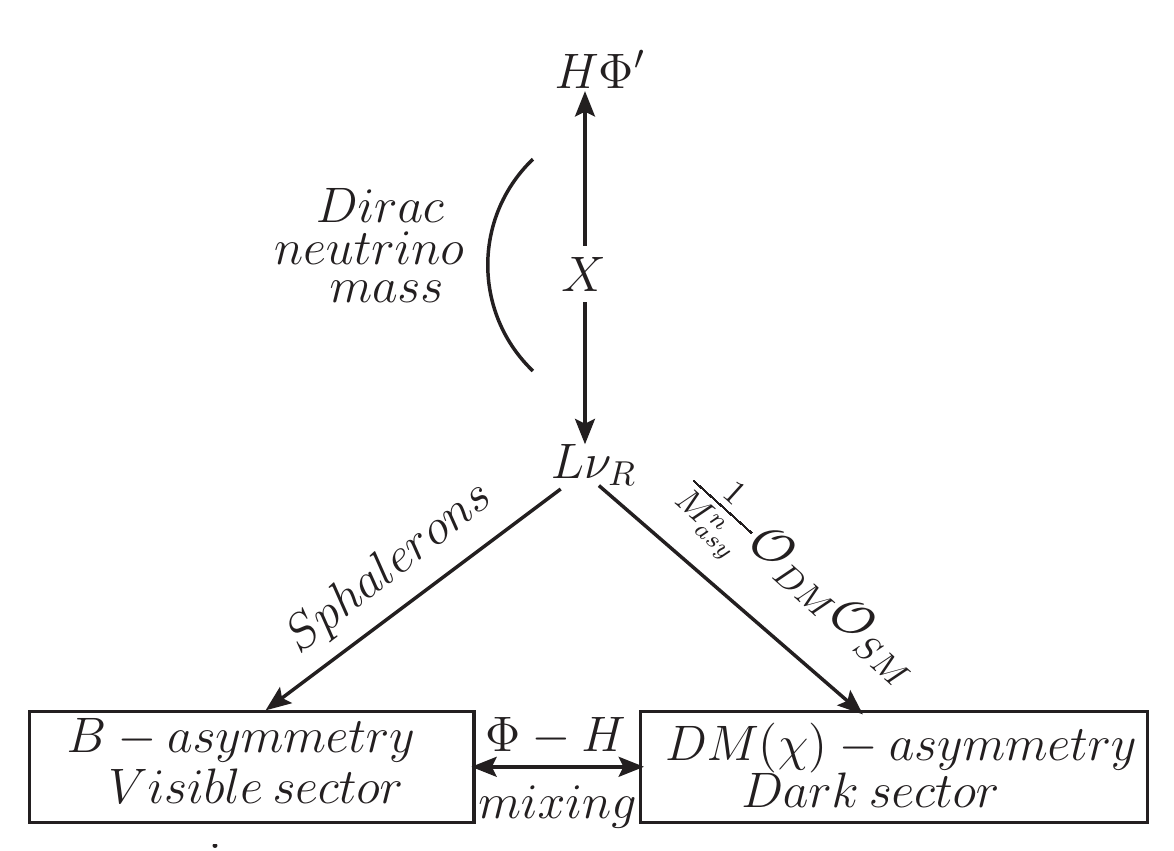}
				 \caption{Pictorial representation shows: i) the generation of neutrino mass, ii) the generation of the $B-L$ asymmetry in the lepton sector. The sphalerons above electroweak (EW) phase transition ($T_{EW}$) transfers a part of $B-L$ asymmetry to the observed B-asymmetry, while the higher dimension operator $\mathcal{O}_{8}$ is in thermal equilibrium above EW-phase transition, at $T > T_{EW}$, cooks a net DM asymmetry out of the existing $B-L$ asymmetry. The $\Phi - H$ mixing provides a bridge between the visible and dark sectors. }
				 \label{diagram}
\end{figure}

Thus, both the SM and the $\Lambda{\rm CDM}$ models are inadequate to explain a plethora of mysteries in physics. At present, the nature of neutrinos, 
either Dirac or Majorana, is not confirmed yet. In future, the neutrinoless double beta decay experiments~\cite{onu_beta_expt} may shed light on it. 
In this paper, assuming neutrinos to be Dirac (i.e. $B-L$ is an exact symmetry) and the DM to be self-interacting, we make an attempt to explain 
simultaneously the three most puzzling physical phenomena {\it viz.} neutrino mass, DM and the baryon asymmetry of the Universe along with a natural 
explanation of the ratio $\Omega_{\rm DM} \approx 5\, \Omega_{\rm B}$. We extend the SM with heavy $SU(2)_L$ 
scalar doublets $X_i$, ($i=1,2$), and $\eta$, singlet right-handed neutrinos $\nu_{iR}, (i=1,2,3)$ and singlet scalars $\Phi$, $\Phi'$ along with 
a vector-like singlet Dirac fermion $\chi$ which represents the SIDM candidate. Since we assume that $B-L$ is an exact symmetry, the CP-violating 
out-of-equilibrium decay of the lightest $X$: {\it i.e.}$X \to H\Phi'$ and $X \to L \nu_{iR}$ generates equal and opposite $B-L$ asymmetry between 
$\nu_{L}$ and $\nu_R$~\cite{Dirac_leptogenesis_1,Dirac_leptogenesis_1,Dirac_leptogenesis_2,Dirac_leptogenesis_3,Dirac_leptogenesis_4,
Dirac_leptogenesis_5,Dirac_leptogenesis_6,Dirac_leptogenesis_7}. After $\Phi'$ and $H$ acquire vacuum expectation values (vev), we get Dirac mass of 
neutrinos of appropriate order. By introducing an additional $U(1)_D$ symmetry (which forbids $\bar{L}\tilde{H} \nu_R$ coupling) we ensure that 
$\nu_L-\nu_R$ equilibration does not occur until below the electroweak (EW) phase transition during which a part of the lepton asymmetry gets converted 
to dark matter asymmetry through a dimension eight operator: $\mathcal{O}_{8}=\frac{1}{M_{asy}^{4}}\bar{\chi}^{2}(LH)^{2}$\cite{Kaplan:2009ag,Feng:2012jn,Ibe:2011hq,Feng:2013wn,Narendra:2018vfw}, $M_{asy}$ being fixed by the mass and relevant couplings of $\eta$. %, where $\chi$ is a vector-like singlet fermion representing dark matter. 
A part of the remaining $B-L$ asymmetry then gets converted to a net B-asymmetry through EW-sphalerons which are active at a temperature above 100 GeV. The singlet 
scalar $\Phi$ not only mediates the self-interaction among the DM particles, but also helps in depleting the symmetric component of the DM. Moreover it mixes 
with the SM Higgs providing a portal for detecting the DM at terrestrial laboratories. We constrain the scalar portal mixing with recent data from experiments 
like CRESST-III and XENON1T. The pictorial presentation can be seen in Fig.\ref{diagram}.
%\vspace*{1cm}

%\vspace{1cm}
%\newpage
The paper is arranged as follows. In section-\ref{Model}, we explain the model for simultaneous explanation of SIDM, baryon asymmetry and Dirac mass of the light neutrinos. In section-\ref{Gen_Asy_DM_transfer}, we explain the Dirac leptogenesis mechanism and production as well as relic density of DM. In section-\ref{sidm}, we find the parameter space for sufficient self-interaction with desired velocity dependence followed by a discussion on the direct detection prospects of the SIDM in section-\ref{sidm_dd}. In section \ref{neutino_mass}, we discuss the parameter space for light neutrino mass and conclude in section-\ref{conclusion}.
% In section-\ref{Gen_Asy_DM_transfer} we find the parameter space for explaining dark matter and baryon asymmetry simultaneously.
%\newpage
%%%%%%%%%%%%%%%%%%%%%%%%%%%%%%%%%%%%%%%%%%%%%%%%%%%%%%%%%%%%%%%%%%%%%%%%%%%%%%%%%%%%%%%%%%%%%%%%%%%
\section{The Model} \label{Model}
%%%%%%%%%%%%%%%%%%%%%%%%%%%%%%%%%%%%%%%%%%%%%%%%%%
We extend the SM, which is based on the gauge group: $ SU(3)_c\times SU(2)_L\times U(1)_Y$, with $U(1)_D\times U(1)_{B-L}$ global symmetries. 
The particle contents of the model are shown in Table~\ref{table} along with their quantum numbers under the imposed symmetry. The scalar sector 
consists of heavy scalar doublets $X_{i}$, ($i=1,2$) and $\eta$, and two singlet scalars $\Phi$, $\Phi'$; while the fermion sector consists of a 
vector-like Dirac fermion $\chi$ along with three heavy right-handed neutrinos $\nu_{R_{i}}, (i=1, 2, 3)$, all singlet under the SM gauge group. 
All these particles carry non-trivial charges under $U(1)_D$ symmetry, while the SM particles remain neutral. Under the extended symmetries, the Majorana 
mass terms of $\chi$ and $\nu_R$ are forbidden.
%
%The $\nu_{R}$ and $\chi$ particles have the same charges under $U(1)_{B-L}$, but they have different charges under $U(1)_{D}$. The 
%scalar doublet $\eta$ also possesses same charge under $U(1)_D$ as that of $\chi$, while the SM particles are neutral under $U(1)_D$ symmetry. Hence 
%both $\chi$ and $\eta$ are part of the dark sector described by $U(1)_D$.  
%When the singlet scalar $\Phi'$ gets a vacuum expectation value (vev), $U(1)_{D}$ symmetry breaks into a remnant $Z_{2}$ symmetry. Under $Z_{2}$ symmetry, $\chi$ and $\eta$ are odd and all other particles are even. 
%
\begin{table}[ht]
\begin{center}
\begin{tabular}{|c|c|c|c|c|}
\hline
Fields & SU(2)$_L$ & U(1)$_Y$ & $U(1)_{D}$ & $U(1)_{B-L}$  \\
\hline

$X_i$ & 2 & +1 & -1 & 0 \\
$\eta$ & 2 & +1 & 1/2 & 0 \\
$\nu_{_R}$ & 1 & 0 & -1 & -1 \\
$\Phi'$ & 1 & 0 & +1 & 0 \\
$\Phi$ & 1 & 0 & 0 & 0 \\
$\chi$ & 1 & 0 & 1/2 & -1 \\
\hline
\end{tabular}
\end{center}
\caption{The additional particle content and their quantum numbers under the imposed symmetry.}
\label{table}
\end{table}

Owing to the symmetry and the charge assignment as shown in Table ~\ref{table}, the Lagrangian of the Model can be written as,\footnote{For simplicity we suppress the indices and state when they require explicitly.}
%\textcolor{red}{
\begin{eqnarray}
-\mathcal{L} &\supset& m_{\chi}\overline{\chi} \chi + \lambda_{_D} \overline{\chi} \chi \Phi + y \,\overline{L}\Tilde{X} \nu_{_{R}} +\rho \,\,\Phi'^{*} X^{\dagger} H  \nonumber\\
&& +  \lambda ~\overline{\chi}\, L \,\eta  + {\rm h.c.}\,
+ V(X, \eta, H, \Phi, \Phi') ,
\label{lag}
\end{eqnarray}
%}
where,
\begin{eqnarray}
V(\eta, H, \Phi, \Phi')&=&M_{\eta}^{2} (\eta^{\dagger} \eta) + \lambda_{\eta}(\eta^{\dagger} \eta)^{2} + \lambda'_{_{\eta H}} (\eta^{\dagger}\eta) (H^{\dagger}H)  \nonumber\\
&& +[\lambda_{_{\eta H}} (\eta^{\dagger} H)^{2} + h.c.]  \nonumber\\
&& -\mu_{H}^{2} H^{\dagger} H + \lambda_{H} (H^{\dagger} H)^{2} + \frac{1}{2} m_{\phi}^{2} \Phi^{2}+ \frac{1}{3} \mu_\Phi \Phi^3 \nonumber\\
&& + \frac{1}{4} \lambda_{\Phi} \Phi^{4} - \mu_{\Phi'}^{2} (\Phi'^{\dagger} \Phi') + \lambda_{\Phi'} (\Phi'^{\dagger} \Phi')^{2}   \nonumber\\
&& + \frac{\mu_{1}}{\sqrt{2}} \Phi H^{\dagger} H + \frac{\mu_{2}}{\sqrt{2}} \Phi (\Phi'^{\dagger} \Phi') + \frac{\lambda_{H \Phi}}{2} H^{\dagger} H \Phi^{2} \nonumber\\
&& + \lambda_{H \Phi'}H^{\dagger} H (\Phi'^{\dagger} \Phi') + \frac{\lambda_{\Phi \Phi'}}{2}\Phi^{2}(\Phi'^{\dagger} \Phi').
\label{potential}
\end{eqnarray}

%%%%%%%%%%%%%%%%%%%%%%%%%%%%%%%%%%%%%%%%%%%%%%%%%%%%%%%%%%%%%%%%%%%%%%%%
The term $\lambda_{_{\eta H}} (\eta^{\dagger} H)^{2} + h.c.$ in the potential (\ref{potential}), breaks the $U(1)_{D}$ symmetry 
softly to a remnant $Z_2$ symmetry under which $\eta$ and $\chi$ are odd, while all other particles are even. We assume that $\eta$ doesn't 
acquire a vev in order to preserve the remnant $Z_2$ symmetry. Considering $M_{\eta} \gg M_{\chi}$, $\chi$ becomes the viable DM candidate. In 
this way, we ensure the stability to DM $\chi$ as well as the theory escapes from having a Goldstone boson.

\begin{figure}[h!]
				\centering
				\includegraphics[width = 56mm]{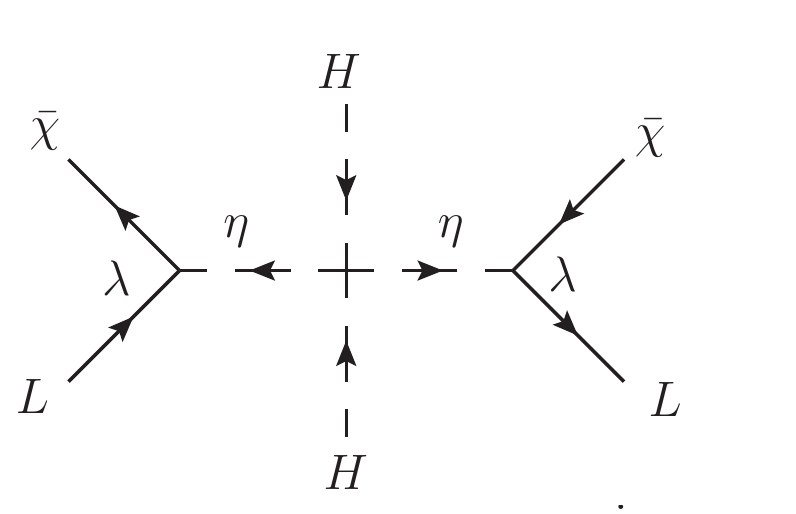}
%				\hfill
				\qquad
				\includegraphics[width = 55mm]{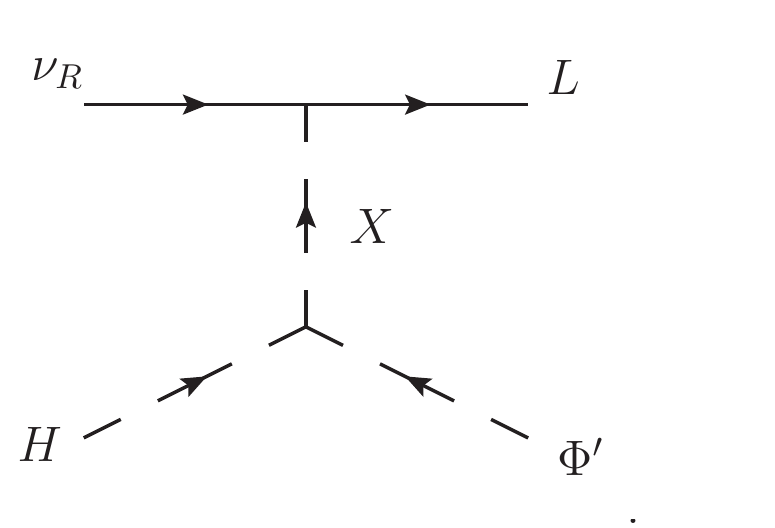}
               \caption{Top: Feynman diagram of the dimension-8 operator. Bottom: Feynman diagram of the dimension-5 operator.}
              \label{Feyn_dim}
\end{figure}

The Lagrangian terms $\lambda ~\overline{\chi} L \eta + \lambda_{_{\eta H}}(\eta^{\dagger} H)^{2} + {\rm h.c.}$ give rise to a dimension-8 transfer 
operator via the Feynman diagram shown in top panel of Fig.~\ref{Feyn_dim}. In the later part of the draft, we will show that this operator transfers 
the lepton asymmetry from visible sector to dark sector. At low energy, upon integrating out the heavy scalar $\eta$, we get the dimension-8 
operator \cite{Kaplan:2009ag,Feng:2012jn,Ibe:2011hq,Feng:2013wn,Narendra:2018vfw}:
\begin{equation}
\mathcal{O}_{8} = \frac{\lambda^{2} \lambda_{_{\eta H}} \overline{\chi} L H H \overline{\chi} L}{M_{\eta}^{4}} \equiv \frac{\overline{\chi}^{2}(L H)^{2}}{M_{asy}^{4}}\,,
\label{eff_O8}
\end{equation}
where $M_{asy}^{4}=M_{\eta}^{4}/(\lambda^{2} \lambda_{_{\eta H}})$. 
Similarly, the Lagrangian terms $y \overline{L}\Tilde{X} \nu_{_{R}} +\rho \,\Phi'^{*} X^{\dagger}H$ lead to an effective dimension-5 operator 
by integrating out the the heavy scalar field $X$ at low energy. The Feynman diagram for this dimension-5 operator $\mathcal{O}_{5}=y \frac{\rho}{M_{X}^{2}} 
\overline{L} H \Phi' \nu_{R}$ is shown in the bottom panel of Fig.~\ref{Feyn_dim}. 

In writing down the scalar potential as given in Eq.\,\ref{potential}, we assume that the scalar doublet $X$ being 
very heavy, do not participate in low energy phenomenology and have implications only in leptogenesis. Here $\mu^2_{H}, m^2_{\phi}, \mu^2_{\Phi'} > 0$. We further assume that $\mu_{\Phi}$ and $\lambda_{\Phi}$ are small. As a result, $\Phi$ acquires a non-zero vev 
through the trilinear terms $\mu_{1}\Phi (H^{\dagger}H)$ and $\mu_{2}\Phi (\Phi'^{\dagger}\Phi')$, which are proportional to the coupling constants $\mu_{1}$ and $\mu_{2}$ respectively. The fluctuations of the fields $\Phi'$, $H$ and $\Phi$ around their corresponding vev's can be written as, 
\begin{equation}
 \Phi'=\frac{w+ \phi'}{\sqrt{2}}
\,\,\,,\,\,\,
H= \begin{bmatrix}  0  \\
   \frac{v+h}{\sqrt{2}}  
   \end{bmatrix}
\,\,\,\,\,{\rm and}\,\,\,\,\,
\Phi= u+\phi  .
\end{equation}

From Eq. \ref{potential}, the stationary conditions are obtained as:
\begin{equation}
\begin{aligned}
w(-\mu^2_{\Phi'}+\lambda_{\Phi'} w^2 + \frac{\mu_2}{\sqrt{2}} u + \frac{\lambda_{H \Phi'}}{2} v^2 + \frac{\lambda_{\Phi \Phi'}}{2} u^2)& = 0 ,
 \\
v(-\mu^2_{H}+\lambda_{H} v^2 + \frac{\mu_{1}}{\sqrt{2}} u + \frac{\lambda_{H \Phi'}}{2} w^2 + \frac{\lambda_{H \Phi}}{2} u^2) &= 0 ,
\\
u(m^2_{\phi} + \mu_\Phi u + \lambda_\Phi u^2 + \frac{\lambda_{H \Phi}}{2} v^2 + \frac{\lambda_{\Phi \Phi'}}{2} w^2) \\ +\frac{\mu_{1}}{2\sqrt{2}}v^2 + \frac{\mu_{2}}{2\sqrt{2}}w^2 = 0 .
\end{aligned}
\label{eq6}
\end{equation}

Assuming $\mu_{\Phi} << m_\Phi \sim 10^{-3}$ GeV and $\lambda_{\Phi} << 1$, we can drop the terms proportional to $\mathcal{O}(u^2)$ and $\mathcal{O}(u^3)$. In this limit, Eq.\ref{eq6} can be rewritten as,
\begin{equation}
\begin{aligned}
w(-\mu^2_{\Phi'}+\lambda_{\Phi'} w^2 + \frac{\mu_2}{\sqrt{2}} u + \frac{\lambda_{H \Phi'}}{2} v^2 ) = 0 ,
 \\
v(-\mu^2_{H}+\lambda_{H} v^2 + \frac{\mu_{1}}{\sqrt{2}} u + \frac{\lambda_{H \Phi'}}{2} w^2) =0 ,
\\
u(m^2_{\phi} + \frac{\lambda_{H \Phi}}{2} 
v^2 + \frac{\lambda_{\Phi \Phi'}}{2} w^2)+\frac{\mu_{1}}{2\sqrt{2}}v^2 + \frac{\mu_{2}}{2\sqrt{2}}w^2=0 .
\end{aligned}
\label{eq7}
\end{equation}

Thus, the vevs of the fields $\Phi'$, $H$ and $\Phi$ are obtained as:
\begin{equation}
\begin{aligned}
 w &= \sqrt{\frac{\mu^2_{\Phi'} - \frac{\mu_{2}}{\sqrt{2}} u - \frac{\lambda_{H \Phi'}}{2} v^2  }{\lambda_{\Phi'}}} ,
\\
v &= \sqrt{\frac{\mu^2_{H} - \frac{\mu_{1}}{\sqrt{2}} u - \frac{\lambda_{H \Phi'}}{2} w^2}{\lambda_H}} ,
\\
u &= \frac{-(\mu_1 v^2 + \mu_2 w^2)}{2\sqrt{2}(m^2_\phi + \frac{\lambda_{H\Phi}}{2} v^2 + \frac{\lambda_{\Phi \Phi'}}{2} w^2)} .
\end{aligned} 
\label{eq8}
\end{equation}

The smallness of $u$ in Eq.~\ref{eq8} can be justified by assuming $\mu_1, \mu_2 \to 0$. The squared mass matrix for the neutral components of three scalar fields $\Phi'$, H and $\Phi$ {\it viz.} $\phi', h, \phi$ can be written in the basis $(\phi', h, \phi)$ as:

%\vspace*{5cm}

%\begin{widetext}
\begin{equation}
 \begin{pmatrix}  
 2\lambda_{\Phi'}w^2    &   \frac{1}{2}\lambda_{H\Phi'}vw    &    \frac{1}{2}(\frac{\mu_{2}w}{\sqrt{2}} + \lambda_{\Phi \Phi'}uw)   \\
\frac{1}{2}(\lambda_{H\Phi'}vw)     &  2 \lambda_{H} v^{2}   &  \frac{1}{2}(\frac{\mu_{1}v}{\sqrt{2}}+\lambda_{H \Phi}vu)  \\ 
 \frac{1}{2}(\frac{\mu_{2}w}{\sqrt{2}} + \lambda_{\Phi \Phi'}uw)    & \frac{1}{2}(\frac{\mu_{1}v}{\sqrt{2}}+\lambda_{H\Phi}vu)     &   m^2_{\phi}+\frac{\lambda_{H\Phi}}{2} v^2+\frac{\lambda_{\Phi\Phi'}}{2}w^2 \\
   \end{pmatrix}.
%\label{scalar_mass_matrix}
\label{matrix_scalar}
\end{equation}
%\end{widetext}
%
As we need light messengers (in MeV scale) to realize sufficiently large DM self-interaction, which also helps in annihilating away the symmetric 
component of DM, we consider the induced vev u of the field $\Phi$ to be very small compared to that of SM Higgs. The details of the diagonalization is given in Appendix~\ref{appendix_diag}. After complete diagonalization, we are left with three mass eigenstates $h_1, h_2$ and $h_3$.  
%$m_{\phi'} \sim m_h \gg m_\phi$. 
The eigenstate $h_1$ with mass $m_{h_1} \approx m_h = 125.18 \,{\rm GeV}$ can be identified as the SM Higgs. The eigenstate $h_2$ with mass $m_{h_2} \sim m_{\Phi'} \sim 10^{5}$ has implications in addressing the neutrino mass and the extra light scalar $h_3$ with mass $m_{h_3} \sim m_\phi$ ($\approx $ MeV scale) mediates the DM self-interaction and annihilates away the symmetric DM component.

%\vspace{1cm}

%%%%%%%%%%%%%%%%%%%%%%%%%%%%%%%%%%%%%%%%%%%%%%%%%%%%%%%%%%%%%%%%%%%%%

\begin{figure}[h!]
		\centering
		\includegraphics[width=3cm,height=2.8cm]{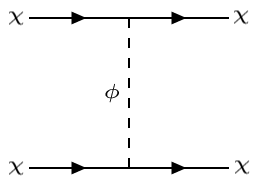}
		\hspace*{2cm}
		%\hfill
		\includegraphics[width=3cm,height=3cm]{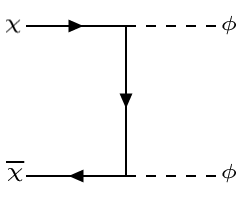}
		\caption{Left: Feynman diagram for elastic DM self-interaction. Right: Dominant annihilation mode for the symmetric component of DM $\chi$.}
		\label{sidm-Feyn}
	\end{figure}

The second term of Eq.\,\ref{lag} gives rise to scalar mediated DM self-interaction via the Feynman diagram shown in the left panel of Fig.~\ref{sidm-Feyn}. The third and fourth terms are responsible for two CP-violating  out-of-equilibrium 
decays: $X \to L \nu_{iR}$ and $X \to H\Phi'$ that generate equal and opposite $B-L$ asymmetry between $\nu_{L}$ and $\nu_R$, required for leptogenesis~\cite{Sakharov:1967dj,sakharov.67,fukugita.86,baryo_lepto_group_1,baryo_lepto_group_2,baryo_lepto_group_3,
baryo_lepto_group_4,baryo_lepto_group_5,baryo_lepto_group_6,baryo_lepto_group_7,baryo_lepto_group_8,baryo_lepto_group_9,Abada:2006ea,Pilaftsis:2003gt,Pilaftsis:2005rv}.	
%The lightest heavy scalar $X$ creates an equal and opposite amount of $B-L$ asymmetry in the visible sector via its CP-violating out-of-equilibrium decays
The dimension-8 operator $\mathcal{O}_{8}$ remains in thermal equilibrium until below the electroweak phase transition, during which it transfers the lepton 
number to the DM number in a $B-L$ conserving way in order to establish a proportionality between DM and baryon densities. The strong Yukawa couplings of 
SM charged leptons rapidly cancel the left and right-handed numbers through LR equilibration processes. But the situation is different for neutrinos due to their tiny Yukawa couplings. They equilibrate after the sphalerons go out of thermal equilibrium. By that time the sphalerons convert a part of left-handed neutrino 
asymmetry to the desired B-asymmetry ~\cite{Dick,Cerdeno:2006ha,Gu:2006dc,Gu:2007mc,Murayama:2002je,Thomas:2005rs,Borah:2016zbd,Perez:2021udy}, while the 
dimension-8 operator $\mathcal{O}_{8}$ keeps on cooking a net DM asymmetry from the existing neutrino asymmetry~\cite{old_DM_Baryon_asy_1, old_DM_Baryon_asy_2, old_DM_Baryon_asy_3, old_DM_Baryon_asy_4, old_DM_Baryon_asy_5,
old_DM_Baryon_asy_6,old_DM_Baryon_asy_7,old_DM_Baryon_asy_8, Asydm_models1_1,Asydm_models1_2,Asydm_models1_3,Asydm_models1_4,
Asydm_models1_5,Asydm_models1_6,Asydm_models1_7,Asydm_models1_8,Asydm_models1_9,Asydm_models1_10, Asydm_models2_1,Asydm_models2_2,
Asydm_models2_3,Asydm_models2_4,Asydm_models2_5,Asydm_models2_6,Asydm_models2_7,Asydm_models2_8,Asydm_models2_9,Asydm_models3_1,
Asydm_models3_2,Asydm_models3_3,Asydm_review_1,Asydm_review_2}. This mechanism directly connects the small Dirac neutrino masses to the DM abundance and the baryon asymmetry of the Universe. 

The scalar $\phi$ is sufficiently light and mediates the self-interaction between the DM particles. Interestingly, the symmetric component of DM depletes via its annihilation into $\phi$ via the Feynman diagram depicted in the right panel of Fig.~\ref{sidm-Feyn}.
 Moreover, it mixes with the SM Higgs and paves a path to detect the DM at direct search experiments. The vev's of singlet scalar $\langle \Phi' \rangle=w$ and SM-Higgs $\langle H \rangle=v$ gives small neutrino masses via higher dimension operator $\mathcal{O}_{5}=y \frac{\rho}{M_{X}^{2}} \overline{L} H \Phi' \nu_{R}$, where the heavy scalar mass suppresses the weak scale. 

%%%%%%%%%%%%%%%%%%%%%%%%%%%%%%%%%%%%%%%%%%%%%%%%%%%%%%%%%%%%%%%%%%%%%%%%%%%%%%%%%%%%%%%%%%%%%%%%%%%%%%%%%%%%%%%%%%%%%%%%%%%%%%%%%%%%%%%%%%%%%%%%%%%%%%%%%%%%%%%%%%%%%%%%%%%%%%%%%%%%%%%%%%%%%%%%%%%%%%%%%%%%%%%%%%%%%%%%%%%%%%%%%%%%%%%%%%%%%%%%%%%%%%%%%%%%%%%%%%%%%%%%
%\begin{figure}[htb!]
%		\includegraphics[width=5cm,height=4cm]{feyn2n.png}
		%\hfill
		%\includegraphics[width=8cm,height=5cm]{thermal_relic.pdf}~
%		\caption{Dominant annihilation mode for the symmetric component of DM $\chi$.}% Right: Under-abundant relic of symmetric component of the DM}
%		\label{fig:feyn2}
%	\end{figure} 	
%%%%%%%%%%%%%%%%%%%%%%%%%%%%%%%%%%%%%%%%%%%%%%%%%%%%%%%%%%%%%%%%%%%%%%%%%%%%%%%%%%%%%%%%%%%%%%%%%%%%%%%%%%%%%%%%%%%%%%%%%%%%%%%%%%%%
%	\section{Direct Detection}\label{sec5}
	%\label{dd}

	%\begin{figure}[h!]
	%	\centering
	%	\includegraphics[width=8cm,height=8cm]{sd_sidm_dd_xenon1t}
		%\hfil
		%\includegraphics[scale=0.45]{rhn_sidm_dd_xenon1t_rep}
	%	\caption{Constraints from DM direct detection in the plane of DM mass $(m_\chi)$ versus mediator mass $(m_\phi)$ for self-interaction.}
	%	\label{sidmdd}
	%\end{figure}
%%%%%%%%%%%%%%%%%%%%%%%%%%%%%%%%%%%%%%%%%%%%%%%%%%%%%%%%%%%%%%%%%%%%%%%%%%%%%%%%%%%%%%%%%%%%%%%%%%%%%%%%%%%%%%%%%%%%%%%%%%%%%%%%%%%%%%%%%%%%%%%%%%%%%%%%%%%%%%%%%%%%%%%%%%%%%%%%%%%%%%%%%%%%%%%%%%%%%%%%%%%%%%%%%%%%%%%%%%%%%%%%%%%%%%%%%%%%%%%%%%%%%%%%%%%%%%%%%%%%%%%%

\section{Dirac Leptogenesis and the DM Relic Density}\label{Gen_Asy_DM_transfer}
%%%%%%%%%%%%%%%%%%%%%%%%%%%%%%%%%%%%%%%%%%%%%%%%%%%%%%%%%%%%%%%%%%%%%%%%%%%%%%%%%%%%%%%%%%%%%%%%%%%
As the Universe cools down, the heavy scalar doublet $X$ which is assumed to have existed in the early Universe, go out of thermal equilibrium below its mass scale and decay in a CP violating way through the channels: $X_{1} \to L \nu_{R}$ and $X_{1} \to H\Phi'$ creating an equal and opposite $B-L$ asymmetry in both left and right-handed sector, see Eq.\,\ref{lag}. The total decay rate of $X_{1}$ is given by: 

\begin{equation}
\begin{aligned}
\Gamma_{X_{1}}&=\frac{1}{8\pi}(y_{1}^{2}+\frac{\rho_{1}^{2}}{M_{X_{1}}^{2}})M_{X_{1}}\\
           &=\frac{1}{8\pi}(y_{1}^{2}+f_{1}^{2})M_{X_{1}}\\
          %&=\frac{y_{1}^2}{16\pi}M_{X_{1}} + \frac{1}{8\pi}f^2 M_{X_{1}}\\
\end{aligned}
\label{decay_rate}
\end{equation}

where $\rho_{1}$ and $y_{1}$ are the trilinear and the Yukawa couplings respectively and in the second line of Eq.\,\ref{decay_rate}, we have defined the dimensionless parameter $f_{1}=\frac{\rho_{1}}{M_{X_{1}}}$, $M_{X_{1}}$ being the mass of the heavy scalar $X_{1}$. To get the adequate lepton asymmetry, the dominant decay channel must be $X_{1} \to L \nu_{R}$, for which the branching ratio $B_L(X_{1} \to L\nu_{iR})$ is given by,

\begin{equation}
\begin{aligned}
B_L(X_{1} \to L\nu_R) &=\frac{(y_{1}^2 M_{X_{1}})/8 \pi}{\Gamma_{X_{1}}} \\
                  &=\frac{1}{1+\frac{f_{1}^2}{y_{1}^2}}\\
\end{aligned}
\end{equation}

We assume $B_L(X_{1} \to L\nu_R) \sim 0.9$ in order to generate the desired lepton asymmetry. Therefore, $f_{1}$ and $y_{1}$ differ roughly by an order magnitude. Since parameters $f_{1}$ and $y_{1}$ are also used in section-\ref{neutino_mass} to explain tiny neutrino mass as $y_{1}\sim \mathcal{O}(10^{-4})$ and $f_{1}\sim \mathcal{O}(10^{-5})$, we stick to region $f_{1},y_{1}\lesssim 10^{-4}$ in this section.
%To get the allowed values  of $y$ and $f$ we plot $\log(f)$ versus $\log (y)$ for
%\begin{figure}[ht]
%				\centering
%				\includegraphics[width = 7cm]{logplot.pdf}
%				 \caption{log(f) vs log(y)}
%				 \label{Log f vs Log y}
%\end{figure}
Demanding $\Gamma_{X_{1}} \lesssim H$ at $T=M_{X_{1}}$, where $H=1.67 g_*^{1/2}T^2/M_{\rm Pl}$ is the Hubble expansion parameter, we get $M_{X_{1}} \lesssim 10^{10}\,{\rm GeV}$ for $f_{1},y_{1}\lesssim 10^{-4}$. For the CP-asymmetry to be non zero, we require atleast two doublet scalars $X_i, i=1,2$. With just one X, the CP-asymmetry will be zero as the imaginary part of the couplings become zero, which can be seen in Eq.11. In presence of these doublet scalars and their interactions, they form a mass matrix $M_{\pm}^{2}$. By diagonalising this mass matrix, we get new mass eigenstates $\zeta_{1}^{\pm}$ and $\zeta_{2}^{\pm}$. See for more details~\cite{Ma:1998dx,Narendra:2017uxl}.

\begin{figure}[h!]
	\centering
	\includegraphics[width = 85mm]{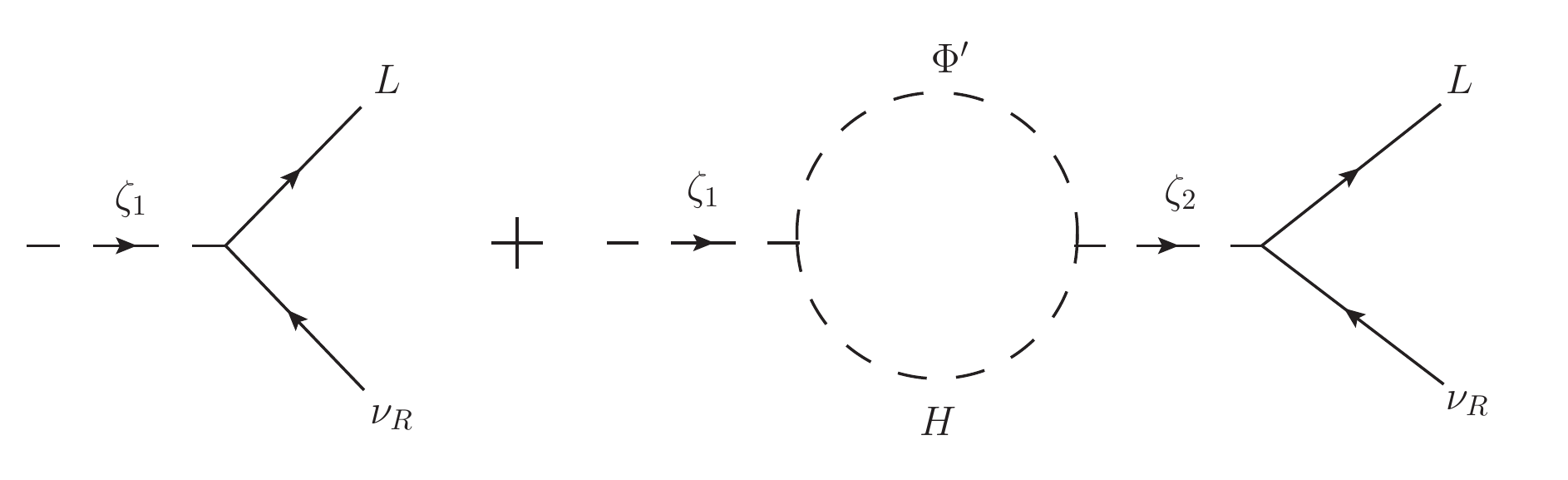}
   \caption{Tree-level and self-energy correction diagrams, whose interference give arise to a net CP-violation in the decay of $\zeta_{1}$.}
       \label{cp_asy_diag}
\end{figure}

We assume a hierarchy among the masses of $\zeta_{1}^{\pm}$ and $\zeta_{2}^{\pm}$ such that the final asymmetry is generated via the 
decay of the lightest one $\zeta_{1}^{\pm}$. The CP-violation arises via the interference of tree-level and one-loop self-energy correction diagrams of lightest 
scalar doublet $\zeta_{1}^{\pm}$ as shown in Fig.~\ref{cp_asy_diag}. 

\begin{figure}[h!]
	\centering
	\includegraphics[width=8cm,height=7.5cm]{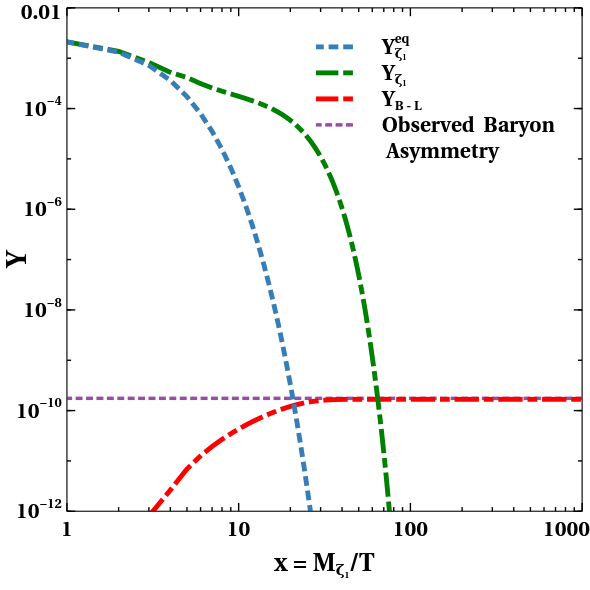}
   \caption{Abundance of $B-L$ and $\zeta_{1}$ as a function of dimensionless variable $x=M_{\zeta_1}/T$. The Red dot-dashed line shows the abundance of $B-L$ asymmetry, where $\epsilon_{L} = 10^{-6}$. The Green dot-dashed line shows the abundance of $\zeta_{1}$. The Blue dotted line shows the equilibrium abundance of $\zeta_{1}$. Here we have taken $y_{1}=10^{-4}$, $\rho_{1} =10^5$ GeV} and $M_{\zeta_{1}}=10^{10}$ GeV.
       \label{yBL_abund}
\end{figure}

The generated asymmetry in the visible sector is then given by
\begin{eqnarray}
\epsilon_{L} &=& [B_L(\zeta_{1}^{-} \rightarrow l^{-} \nu_{R})-B_L(\zeta_{1}^{+}\rightarrow (l^{-})^{c} \nu_{R}^{c})]\nonumber\\
  &=& -\frac{{\rm Im}\left(\rho_{1}^{*} \rho_{2} \mathop{\sum_{k,l}} y^{*}_{1kl} y_{2kl} \right)}{8\pi^2(M_{\zeta_2}^{2}-M_{\zeta_1}^{2})} \left[\frac{M_{\zeta_1}}{\Gamma_{\zeta_1}} \right]\,,
 \end{eqnarray}
where $B_L$ is the branching ratio for $\zeta_{1}^{\pm} \rightarrow l^{\pm} \nu_{R}$ and $M_{\zeta_i}, i=1,2$ are the masses of heavy doublet scalars. 
Here $\rho_{1,2}$ are respectively the mass dimension coupling of $X_{1,2}$ with $\Phi' H$, while $y_{1,2}$ are the respective Yukawa couplings of 
$X_{1,2}$ with $L \nu_R$. We get a net $B-L$ asymmetry~\cite{Narendra:2018vfw, buchmuller&plumacher, Buchmuller:2004nz,Giudice:2003jh}.

\begin{equation}
(n_{B-L})_{total} = \epsilon_{L} \kappa s \times \frac{n_{\zeta_{1}}^{eq} (T \rightarrow \infty )}{s}
\label{nb-ltotal_final}
\end{equation}
where $(n_{\zeta_{1}}^{eq}/s)(T \rightarrow \infty)=135\, \zeta(3)/(4 \pi^{4} g_{*})$ is the relativistic equilibrium abundance of $\zeta_{1}^\pm$, where $\zeta(3)=1.202$. $s=(2\pi^2/45)g_* T^3$ is the entropy density of the co-moving volume. $\kappa$ is the washout factor, which arises due to inverse decay and scattering processes. It can vary between 0 to 1 depending on the strength of the Yukawa coupling. For definiteness we choose $\kappa = 0.01$. The $B-L$ asymmetry in the visible sector can be generated by solving the relevant coupled Boltzmann equations given by Eq.\ref{boltzmann} \cite{Narendra:2017uxl}. In Fig.\ref{yBL_abund}, we show the evolution of the number density of $\zeta_1$, {\it i.e.,} $ Y_{\zeta_1}$ and the $B-L$ asymmetry $Y_{\rm B-L}$ by solving the Boltzmann equations as given below:
\begin{equation}
\begin{aligned}
\frac{dY_{\zeta_1}}{dx} &= - \frac{x}{H(M_{\zeta_1})} s \langle \sigma|v|_{(\zeta_1 \zeta_1 \rightarrow All)} \rangle \Big[Y^2_{\zeta_1} - (Y^{eq}_{\zeta_1})^2\Big] \\
& - \frac{x}{H(M_{\zeta_1})}\Gamma_{(\zeta_1 \rightarrow All)} \Big[Y_{\zeta_1} - Y^{eq}_{\zeta_1}\Big]
\\
\frac{dY_{\rm B-L}}{dx} &= \frac{x}{H(M_{\zeta_1})} \Big[ \epsilon_L \Gamma_{(\zeta_1 \rightarrow All)} B_L \left( Y_{\zeta_1} - Y^{eq}_{\zeta_1} \right) 
-\Gamma_{W} Y_{\rm B-L} \Big],\\
\end{aligned}
\label{boltzmann}
\end{equation}
where $\Gamma_W$ in the second equation takes care of the washout effects corresponding to inverse decay and scattering. The decay 
and the inverse decay of $\zeta_1^\pm$ are related by $\Gamma_{inv}=(Y_{\zeta_1}^{eq}/Y_{\ell}^{eq}) \Gamma_{(\zeta_1 \rightarrow All)}$. The inverse 
decay of $\zeta_1^\pm$ falls exponentially after the latter goes out-of-equilibrium. Similarly the lepton number conserving 
% \cite{Buchmuller:2000as}. 
$2\rightarrow 2$ scattering processes: $\nu_R \Phi' \to L H$, $\nu_R H \to L \Phi'$, $\nu_R \overline{L} \to H \Phi'$ mediated by $X$-particles are 
suppressed due to the small Yukawa couplings required for generating light neutrino masses of Dirac type in section-VI. The resulting $B-L$ asymmetry 
is shown by the red dot-dashed line in Fig. 5.
%\cite{Cerdeno:2006ha}.}
%Fig.\ref{yBL_abund} shows the abundance of $B-L$ and $\zeta_{1}$ as function of dimensionless variable $x=M_{1}/T$.

In the following we discuss the distribution of the generated $B-L$ asymmetry between the visible sector and the dark sector. Once 
the $B-L$ asymmetry is generated in the visible sector, the dimension-8 operator $\mathcal {O}_8= \frac{1}{M_{asy}^{4}} \overline{\chi}^{2}(LH)^{2}$ 
transfers it partially to the dark sector~\cite{Feng:2012jn}. This requires the $\mathcal{O}_{8}$ operator to be in thermal equilibrium above the 
sphaleron decoupling temperature $T_{\rm sph}$. For Higgs mass $M_{h_1}=125$ GeV, $T_{\rm sph}\gtrsim M_W$, where $M_W$ is the W-boson mass. As 
discussed in refs. \cite{Narendra:2018vfw, Narendra:2019cyt}, the $\mathcal{O}_{8}$ operator remains in thermal equilibrium down to a temperature 
$T_D \gtrsim T_{\rm sph} = M_W$ for $M_{asy} \gtrsim 0.9 \times 10^{4}$ GeV. As a result the DM remains in thermal contact with the visible 
sector via this $\mathcal{O}_{8}$ operator down to a decoupling temperature $T_D \gtrsim T_{\rm sph} = M_W$, during which a net $B-L$ asymmetry 
of the $\chi$-particles is generated.The details of the chemical equilibrium calculation are given in appendix~\ref{chemical}. 
The number density of $\chi$ is then given by:
\begin{equation}
n_\chi = (n_{\rm B-L})_{\rm dark} = \frac{58}{291}(n_{B-L})_{\rm vis}\,\,.
\label{a1}
\end{equation}
The total $B-L$ asymmetry created in the early universe is through out-of-equilibrium decays of $\zeta_{1}$ and it is the only source of $B-L$ asymmetry in both dark and visible sectors. So we have the following condition,
\begin{eqnarray}
(n_{B-L})_{\rm total} &=& (n_{B-L})_{\rm vis} + (n_{B-L})_{\rm dark}\nonumber\\
&=& \frac{349}{291}(n_{B-L})_{\rm vis}.
\label{a2}
\end{eqnarray}
Comparing Eq.\,\ref{a2} with Eq.\,\ref{nb-ltotal_final} and using $n_{\rm B}=0.31\,(n_{\rm B-L})_{\rm vis}$~\cite{Narendra:2018vfw}, we get the required CP asymmetry for observed lepton abundance to be $\epsilon_{L} = 141.23 (\eta/\kappa)$.  For $\kappa \sim 0.01$, the required CP-asymmetry is $\epsilon_{L} \sim 10^{-6}$. 
Using Eq.\,\ref{a2} in $n_{\rm B}=0.31\,(n_{\rm B-L})_{\rm vis}$ and Eq.\,\ref{a1}, we get,
\begin{equation}
n_{\rm B}=\frac{90}{349} (n_{\rm B-L})_{\rm total} \,\,\,\,,\,\,\,\, n_\chi = \frac{58}{349}(n_{B-L})_{\rm total}
\label{nchi_nB_total}
\end{equation}
Therefore,
\begin{equation}
\frac{n_{\chi}}{n_{B}}=\frac{58}{90}.
\label{asy_1}
\end{equation}

The expression for the relic of the asymmetric dark matter component is given by, 
\begin{equation}
\Omega_{asy}=n_{\chi}m_{\chi}/\rho_{c}\,\,,
\label{asy_2}
\end{equation}
where $\rho_{c}$ is the critical density. For fully asymmetric dark matter $\Omega_{DM}=\Omega_{asy}$, so we have,
\begin{equation}
\frac{\Omega_{DM}}{\Omega_{B}}=\frac{\Omega_{asy}}{\Omega_{B}}=\frac{n_{\chi}m_{\chi}}{m_{B}n_{B}}=\frac{58}{90}m_{\chi}.
\label{asy_3}
\end{equation}

From WMAP and Planck data, we have the ratio between dark matter density to baryonic matter density,
\begin{equation}
 \frac{\Omega_{DM}}{\Omega_{B}}\sim 5.
 \label{asy_4}
\end{equation} 
Using Eq.\,\ref{asy_3} and Eq.\,\ref{asy_4}, we can get the dark matter mass for fully asymmetric dark matter to be
$m_{\chi}=7.76\rm\, GeV$.\\
Now, if we consider that the dark matter is not fully asymmetric, instead it contains both symmetric and asymmetric components,
\begin{equation}
\Omega_{DM}h^2 = (\Omega_{sym}+\Omega_{asy})h^2=0.12.
 \label{asy_5}
\end{equation}

Therefore the percentage of the dark matter relic contributed by asymmetric part is given by,
\begin{equation}
\text{ADM}(\%)=\frac{\Omega_{asy}}{\Omega_{DM}}\times 100=\frac{\Omega_{asy}}{\Omega_{sym}+\Omega_{asy}}\times 100.
 \label{asy_6}
\end{equation} 

Using Eq.\,\ref{asy_4} and Eq.\,\ref{asy_6}, we have,
\begin{equation}
\frac{\Omega_{asy}}{\Omega_{B}}=5\times \frac{\Omega_{asy}}{\Omega_{sym}+\Omega_{asy}} = 5\times \frac{\text{ADM}(\%)}{100}.
\end{equation}

Using Eq.\,\ref{asy_3} and Eq.\,\ref{asy_6}, we can write the dark matter mass for general scenario, where both symmetric and asymmetric component contributing to the dark matter relics as,
\begin{equation}
m_{\chi}=\frac{90}{58}\times 5\times \frac{\text{ADM}(\%)}{100}.
\end{equation}
So depending on how much asymmetric relic contributes to the total DM relic, the mass of the dark matter will change accordingly from the above equation.

\begin{figure}[ht!]
				%\centering
				\includegraphics[width=8cm,height=8cm]{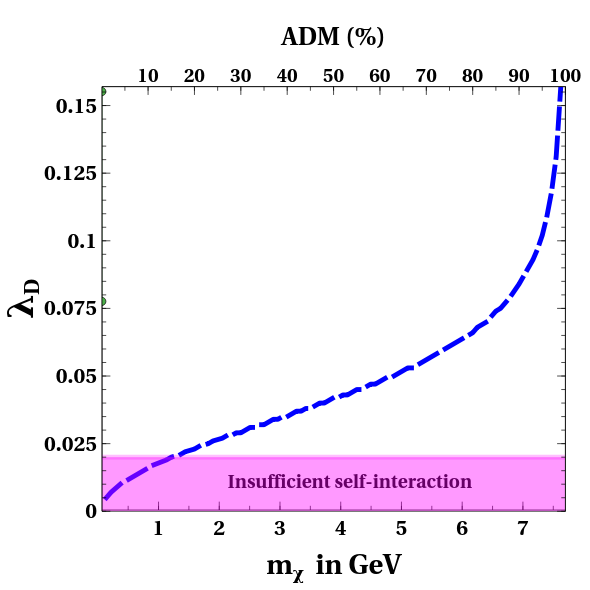}
				\hfill
				\includegraphics[width=7.5cm,height=7.2cm]{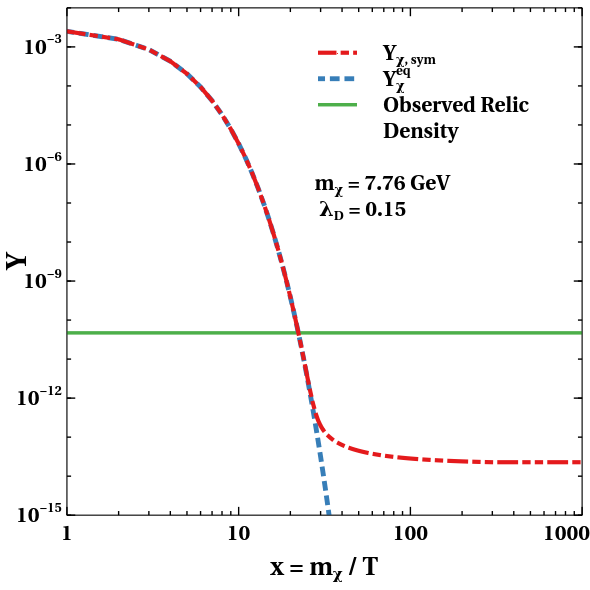}~
				 \caption{Left: $\lambda_{D}$ vs ADM($\%$) plot with corresponding DM mass in order to satisfy the correct relic density. The magenta shaded region is excluded due to insufficient DM self-interaction. Right: Under-abundant relic of symmetric component of the DM as depicted by the dot-dashed red curve. The equilibrium number density is depicted by the dotted blue curve.}
                   \label{asy_perc}             
\end{figure}

In this general scenario, the symmetric component contributing to the total relic is $(100-\text{ADM})\%$. So we have,
%\begin{equation}
\begin{align}
100-\text{ADM}(\%)&=\frac{\Omega_{sym}\times 100}{\Omega_{sym}+\Omega_{asy}} \nonumber \\ 
&=\frac{\Omega_{sym}}{\Omega_{DM}}\times 100=\frac{\Omega_{sym}h^2}{0.12}\times 100, \nonumber \\ 
\Rightarrow \Omega_{sym}h^2 &= \frac{100-\text{ADM}(\%)}{100}\times 0.12 \nonumber \\.
\label{asy_10}
\end{align}
The relics of the symmetric component is determined by the thermal average of its annihilation cross-section, given by,
\begin{equation}
\Omega_{sym} h^2=\frac{8.7661\times 10^{-11}}{\sqrt{g_{*s}}J},
\label{asy_11}
\end{equation}

where $J$ is given by,
\begin{equation}
J= \int_{x_{f}}^{\infty}\frac{\langle\sigma v\rangle_{\overline{\chi}\chi \to~ \phi \phi}}{x^2} \,dx \ .
\label{asy_12}
\end{equation}
where$ \langle\sigma v\rangle_{\overline{\chi}\chi \to~ \phi \phi}$ is the thermally averaged cross-section of the dominant annihilation process $\overline{\chi}\chi \to \phi \phi$ shown in the right panel of Fig.~\ref{sidm-Feyn}, which can be roughly estimated to be,
	\begin{equation}
		\langle\sigma v\rangle_{\overline{\chi}\chi \to~ \phi \phi} \approx	\frac{3}{4}\frac{\lambda^4_D}{16\pi m^2_{\chi}}
		\label{ann}
	\end{equation}
The amount of annihilation can be very large or small depending on the value of $\lambda_D$. This fixes the contribution of the symmetric (and hence the asymmetric) component to the total DM relic density which in turn fixes the DM mass.  Using Eq.\ref{asy_10}, Eq.\ref{asy_11}, Eq.\ref{asy_12}, and Eq.\ref{ann}, we can determine the $\lambda_{D}$ parameter with respect to ADM percentage (or equivalently DM mass), which is depicted by the dashed blue curve in the left panel of Fig.~\ref{asy_perc}. Considering ADM component to vary within $1-99.99\%$, the parameter $\lambda_{D}$ varies within $0.005-0.15$ and correspondingly allowed DM mass range is $0.07-7.76$ GeV.   

For fully asymmetric dark matter, i.e., $\rm ADM(\%) > 99.99\%$, the relics of the symmetric component should be very small, which requires large $\lambda_{D}$ ($\sim 0.15 $) for efficient annihilation of the symmetric component. Such a large $\lambda_{D}$ in turn give rise to sufficient self-interaction among the DM particles mediated by the light scalar $\phi$ as depicted in the left panel of Fig.~\ref{sidm-Feyn}. We rule out the region $\lambda_D < 0.02$ (correspondingly $m_\chi < 1.3$ GeV) in the top panel of Fig.~\ref{asy_perc} due to insufficient self-interaction ($\sigma/m_{{\rm DM}} < 10^{-24} {\rm cm}^2/{\rm GeV}$) which can not alleviate the small scale anomalies of $\Lambda{\rm CDM}$. We elaborate the details of DM self-interaction in section-\ref{sidm}. For a fully ADM scenario, the evolution of the relic abundance of the symmetric component is shown in the right panel of Fig.~\ref{asy_perc} by solving the relevant Boltzmann equation for the comoving number density of the symmetric component $Y_{\chi,sym} = \frac{n_{\chi,sym}}{s}$, where $n_{\chi,sym}$ is the actual number density and $s$ is the entropy density. The Boltzmann equation is given by,

\begin{equation}
\label{be}
\frac{dY_{\chi,{\rm sym}}}{dx} = -\frac{s(m_\chi)}{H(m_\chi)} \langle\sigma v\rangle_{\overline{\chi}\chi \to~ \phi \phi} \left(Y^2_{\chi,{\rm sym}} - (Y^{eq}_{\chi})^2\right)
\end{equation}

where $s(m_\chi)$ and $H(m_\chi)$ are the entropy density and the Hubble parameter as a function of DM mass defined as,
$$ 
s(m_\chi)= \frac{2 \pi^2 }{45} g_*\, m_\chi^3, \quad H(m_\chi)= \frac{\pi}{\sqrt{90}} \frac{\sqrt{g_*}}{M^r_{pl}} m_\chi^2, $$  where $ M^r_{pl}= 2.44\times {10}^{18} {\rm GeV} $ is the reduced Planck mass and $Y^{\text{eq}}_{\chi}$ is the equilibrium DM number density.

\section{Dark Matter Self-interaction}\label{sidm}
	The DM $\chi$ has elastic self-scattering, mediated by the light scalar $\phi$ as depicted by the Feynman diagram shown in left panel of Fig.~\ref{sidm-Feyn}, thanks to the interaction term $\lambda_D \overline{\chi}\chi \phi$ in the model Lagrangian given by Eq.\,\ref{lag}. % The Feynman diagram of such process is shown in 
	%In order to alleviate the small-scale anomalies of $\Lambda{\rm CDM}$, the typical DM self-scattering cross-section should be $\sigma \sim 1~{\rm cm}^2 /{\rm g} \approx 2\times 10^{-24}~{\rm cm}^2 /{\rm GeV}$, which is 14 orders of magnitude larger than the typical WIMP cross-section($\sigma \sim 10^{-38} \, {\rm cm}^2/GeV$). This suggests the existence of a light mediator, which is much lighter than electroweak scale. The scalar mediator $S$ in our model serves this purpose. 
	The scattering in non-relativistic limit is well-described by the attractive Yukawa potential,
	\begin{equation}
		V(r)= \frac{\lambda^2_D}{4\pi r}e^{-m_{\phi}r}
	\end{equation} 
To capture the relevant physics of forward scattering divergence, we define the transfer cross-section $\sigma_T$ as~\cite{Feng:2009hw,Tulin:2013teo,Tulin:2017ara}
	\begin{equation}
		\sigma_T = \int d\Omega (1-\cos\theta) \frac{d\sigma}{d\Omega}
	\end{equation}
 Depending on the masses of DM ($m_\chi$) and the mediator ($m_{\phi}$), as well as the relative velocity of the colliding particle ($v$) and the coupling ($\lambda^2_D$), we can identify three distinct regimes. The Born regime ($\lambda^2_D m_\chi/(4\pi m_\phi) \ll 1,  m_\chi v/m_{\phi} \geq 1$) is where the perturbative calculation holds good. Outside the Born regime, we have the classical regime ($\lambda^2_D m_\chi/(4\pi m_\phi) \geq 1, m_\chi v/m_{\phi} \geq 1$) and the resonant regime ($\lambda^2_D m_\chi/(4\pi m_\phi) \geq 1, m_\chi v/m_{\phi} \leq 1$) where non-perturbative and quantum-mechanical effects become important. The self-interaction cross-sections in these regimes are listed in Appendix~\ref{appendix2}. In the left panel of Fig.~\ref{sidm1}, we show the self-interaction allowed parameter space in $m_\chi$ - $m_{\phi}$ plane obtained by constraining $\sigma/m_\chi$ in the correct ballpark from astrophysical data across different scales.  We constrain $\sigma/m_\chi$ in the range $0.1-100~{\rm cm}^2/{\rm g}$ for dwarf galaxies ($v\sim 10~ \rm km/s$) as shown by the three shades of blue coloured region as indicated in the figure inset. The light magenta coloured region depicts the parameter space allowed for galaxies ($\sigma/m_\chi \sim 0.1-10~{\rm cm}^2/{\rm g}$), while the green coloured region depicts the parameter space allowed for clusters ($\sigma/m_\chi \sim 0.1-1~{\rm cm}^2/{\rm g}$). The masses of DM and the mediator for which all three regions overlap will alleviate the small scale anomalies across all scales.	
	\begin{figure}[h!]
	\centering
		\includegraphics[width=7.5cm,height=7cm]{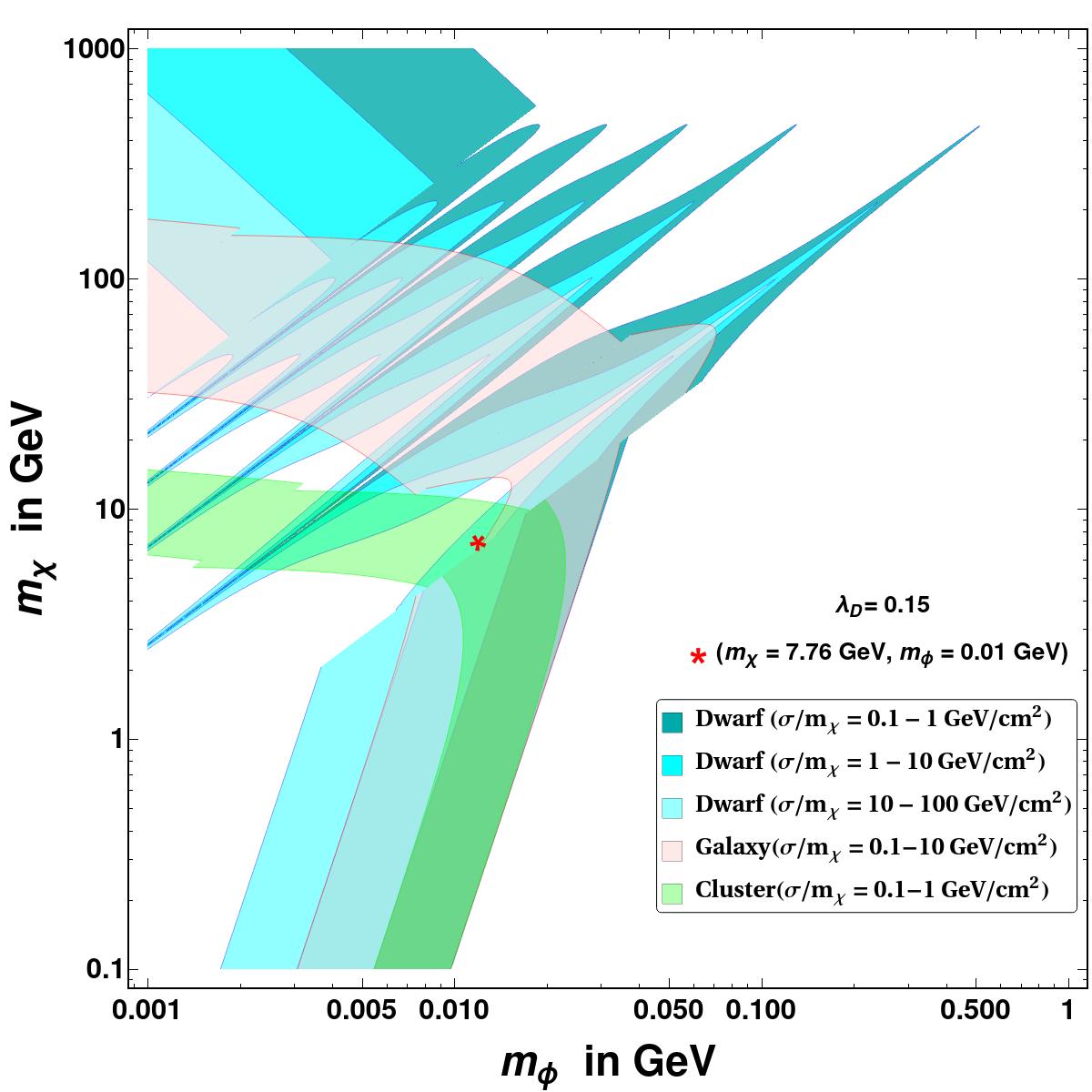}
		\hfill
		\vspace*{1cm}
		\includegraphics[width=7.5cm,height=9cm]{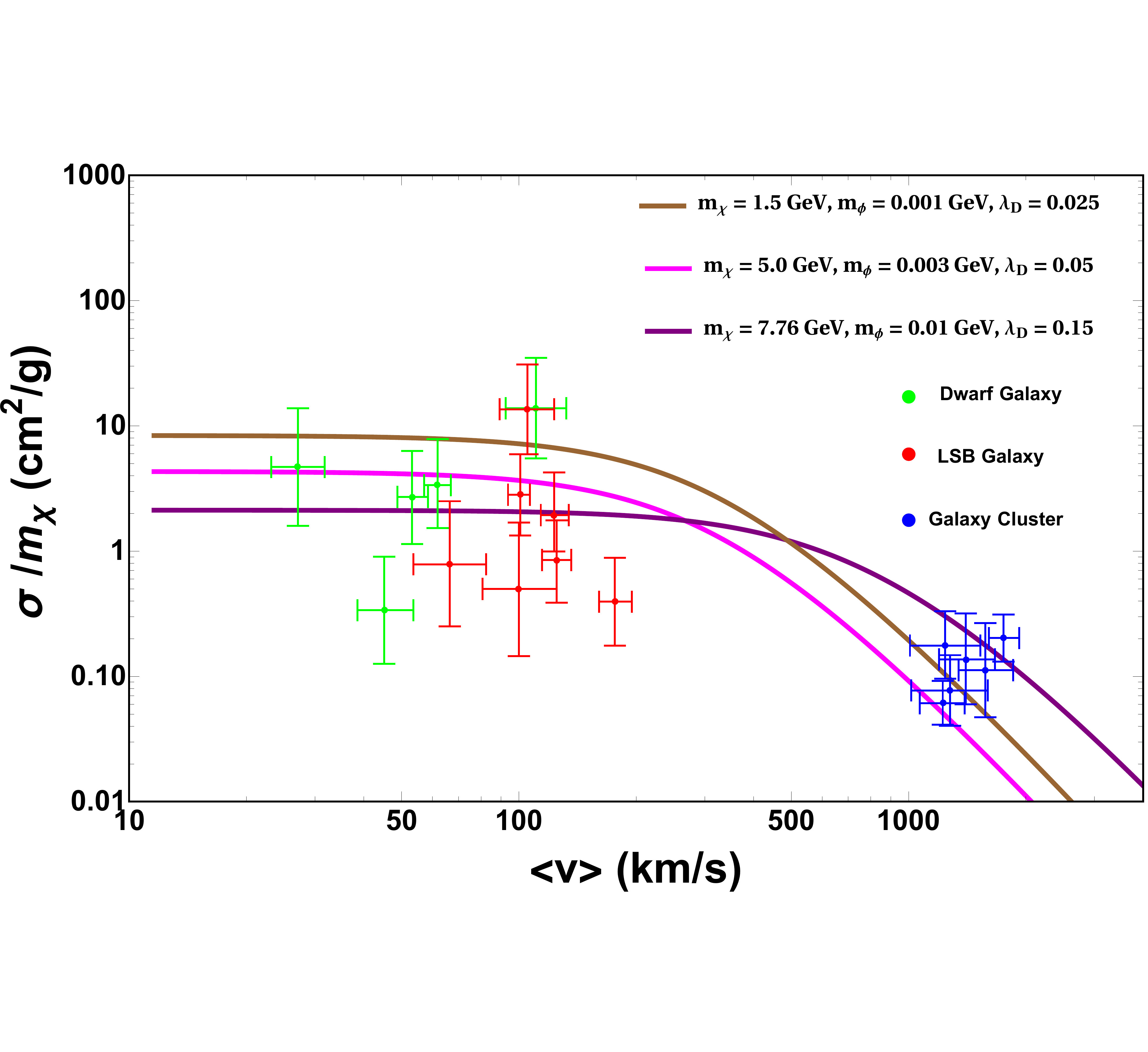}
		\caption{Left: Self-interaction cross-section in the range $0.1-1 \; {\rm cm}^2/{\rm g}$ for clusters ($v\sim1000 \; {\rm km/s}$), $0.1-10 \; {\rm cm}^2/{\rm g}$ for galaxies ($v \sim 200 $ km/s.) and $0.1-1000 \; {\rm cm}^2/{\rm g}$ for dwarfs ($v \sim 10 $ km/s). Right: The self-interaction cross-section per unit mass of DM as a function of average collision velocity.}
		\label{sidm1}
	\end{figure}	
	The top (bottom) corner corresponds to the Classical (Born) region, where the cross-section depends on velocity trivially. The sandwiched region between these two is the so-called resonant region, where quantum mechanical resonances and anti-resonances appear due to (quasi)bound state formation in the attractive potential. The resonances and anti-resonances are most prominent at the dwarf scale due to the lower velocity of the DM particles and gradually becomes less prominent towards galaxy and cluster scales as DM velocity increases and less likely to be bounded. For a coupling $\lambda_D$, the condition $m_\chi v/m_\phi < 1$ dictates the onset of non-perturbative quantum mechanical effects, which is easily satisfied by smaller velocities. We have considered in the left panel of Fig.~\ref{sidm1}, $\lambda_D = 0.15 $ which corresponds to fully asymmetric DM and also is just at the correct ballpark for sufficient self-interaction allowing maximum parameter space in the $m_\chi-m_\phi$ plane. For lower values of $\lambda_D$ (which means DM is not fully asymmetric), the parameter space where desired cross-sections can be obtained gradually decreases. For $\lambda_D < 0.02$, the obtained cross-sections are below the ballpark of $\sigma/m_\chi \sim 0.1 cm^2/g$, insufficient to alleviate the small scale $\Lambda{\rm CDM}$ anomalies.     
	%The resonant spikes are not distinct in these figures as we have varied the Yukawa coupling in a range 0.1-1. Nevertheless, prominent resonant spikes can be seen in Fig.~\ref{sidmdd} in section~\ref{sec5}, where we show the same parameter space for a fixed Yukawa coupling $y'_1=0.35$, while confronting the SIDM parameter space to direct search. We can see from the figures that a wide range of DM mass can give rise to sufficient self-interaction. However, the mass of the mediator is constrained roughly within two orders of magnitudes excepting for the resonance case. 
	%We will confront these regions of parameter space to direct detection bounds in section~\ref{sec5}.
	
	The self-scattering cross-section per unit DM mass as a function of average collision velocity obtained from the model fits to data from dwarfs (red), low surface brightness (LSB) galaxies (blue), and clusters (green)~\cite{Kaplinghat:2015aga, Kamada:2020buc} as shown in the bottom panel of Fig \ref{sidm1}. The purple curve corresponds to the benchmark point of fully asymmetric DM ($m_{\chi}=7.76~\rm GeV$) with $m_\phi=0.01 {\rm GeV}$ and $\lambda_D = 0.15 $. We depict this point with a red star mark in Fig.~\ref{sidm1} as well. The brown curve corresponds to the benchmark point of $m_{\chi}=1.5~\rm GeV$ with $m_\phi=0.001~ {\rm GeV}$ and $\lambda_D = 0.025$ which just cuts the mark to give sufficient self-interaction as shown in the top panel of Fig.~\ref{asy_perc}. The magenta curve corresponds to an intermediate case with $m_{\chi}=5~\rm GeV$, $m_\phi=0.003~ {\rm GeV}$ and $\lambda_D = 0.05$. Hence, it is clear from the bottom panel of Fig.~\ref{sidm1} that the model can appreciably explain the astrophysical observation of velocity-dependent DM self-interaction.	
	%\begin{figure}
	%	\includegraphics[width=8cm,height=9cm]{astro_fit.png}
	%	\caption{The self-interaction cross section per unit mass of DM as a function of average collision velocity.}
	%	\label{astrofit}
	%\end{figure} 
%The annihilation cross-section mainly depends on the free parameters: $\Phi-H$ nixing, {\it i.e.} $\sin \beta$, the mass of $h_{2}${\it i.e.} $M_{h_{2}}$ and the $\chi$ coupling with $h_{2}$, {\it i.e.} $\lambda_{DM}$. These free parameters are highly constrained by invisible Higgs decay~\cite{Khachatryan:2016whc}, relic density of DM reported by WMAP~\cite{Hinshaw:2012aka} and Planck~\cite{Akrami:2018vks}, SI direct detection cross-sections at XENON100~\cite{Aprile:2012nq}, LUX~\cite{Akerib:2016vxi}, XENON1T~\cite{Aprile:2015uzo}\, and the Higgs signal strength measured at LHC~\cite{cms_report_2018, Khachatryan:2016vau}.
\section{DM Direct Search} \label{sidm_dd}
\begin{figure}[h!]
		\centering
		\includegraphics[scale=0.2]{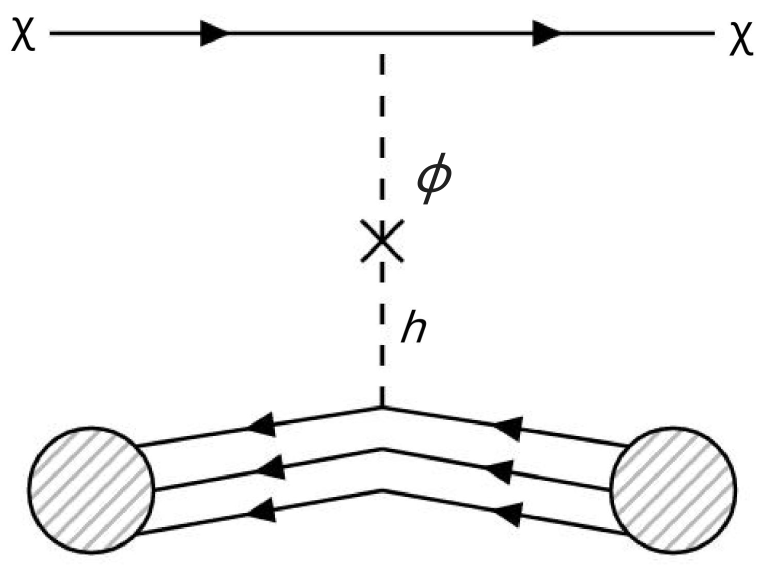}
		\caption{Spin-independent DM-nucleon scattering via scalar mixing.}
		\label{DM_diag}
	\end{figure}

The SIDM can be detected at terrestrial laboratories through $\phi-h$ mixing ($\theta_{\phi h}$), via its scattering off the target nuclei as depicted by the Feynman diagram shown in Fig.\,\ref{DM_diag}. the scattering cross-section of DM per nucleon can be expressed as
	\begin{equation}
		\sigma_{SI}^{\phi-h} = \frac{\mu_{r}^{2}}{4\pi A^{2}} \left[ Z f_{p} + (A-Z) f_{n} \right]^{2}
		\label{DD_cs}
	\end{equation}
	where $\mu_{r}=\frac{m_{\chi}m_n}{m_{\chi} + m_n}$ is the reduced mass of the DM-nucleon system. Here $m_{n}$ is the nucleon (proton or neutron) mass, A and Z are respectively the mass and atomic number of the target nucleus, $f_{p}$ and $f_{n}$ are the interaction strengths of proton and neutron with DM respectively, given as:
	\begin{equation}
		f_{p,n}=\sum\limits_{q=u,d,s} f_{T_{q}}^{p,n} \alpha_{q}\frac{m_{p,n}}{m_{q}} + \frac{2}{27} f_{TG}^{p,n}\sum\limits_{q=c,t,b}\alpha_{q} 
		\frac{m_{p,n}}{m_{q}}\,,
		\label{fpn}
	\end{equation}
	where 
	\begin{equation}
		\alpha_{q} =  \lambda_D \theta_{\phi h}\left( \frac{m_{q}}{v}\right) \left[\frac{1}{m^2_\phi}-\frac{1}{m^{2}_h}\right] \,.
		\label{DD4}
	\end{equation}
	In Eq.\,\eqref{fpn}, the values of $f_{T_{q}}^{p,n}$ can be found in~\cite{Ellis:2000ds}. 
	%The mixing angle $\theta_{\phi h}$ can be derived in terms of the parameters $\lambda_{\phi h},\langle \phi \rangle, v, m_\phi, m_h$.

	\vspace{0.3cm}

\begin{figure}[h!]
		\centering
		\includegraphics[width=8cm,height=8cm]{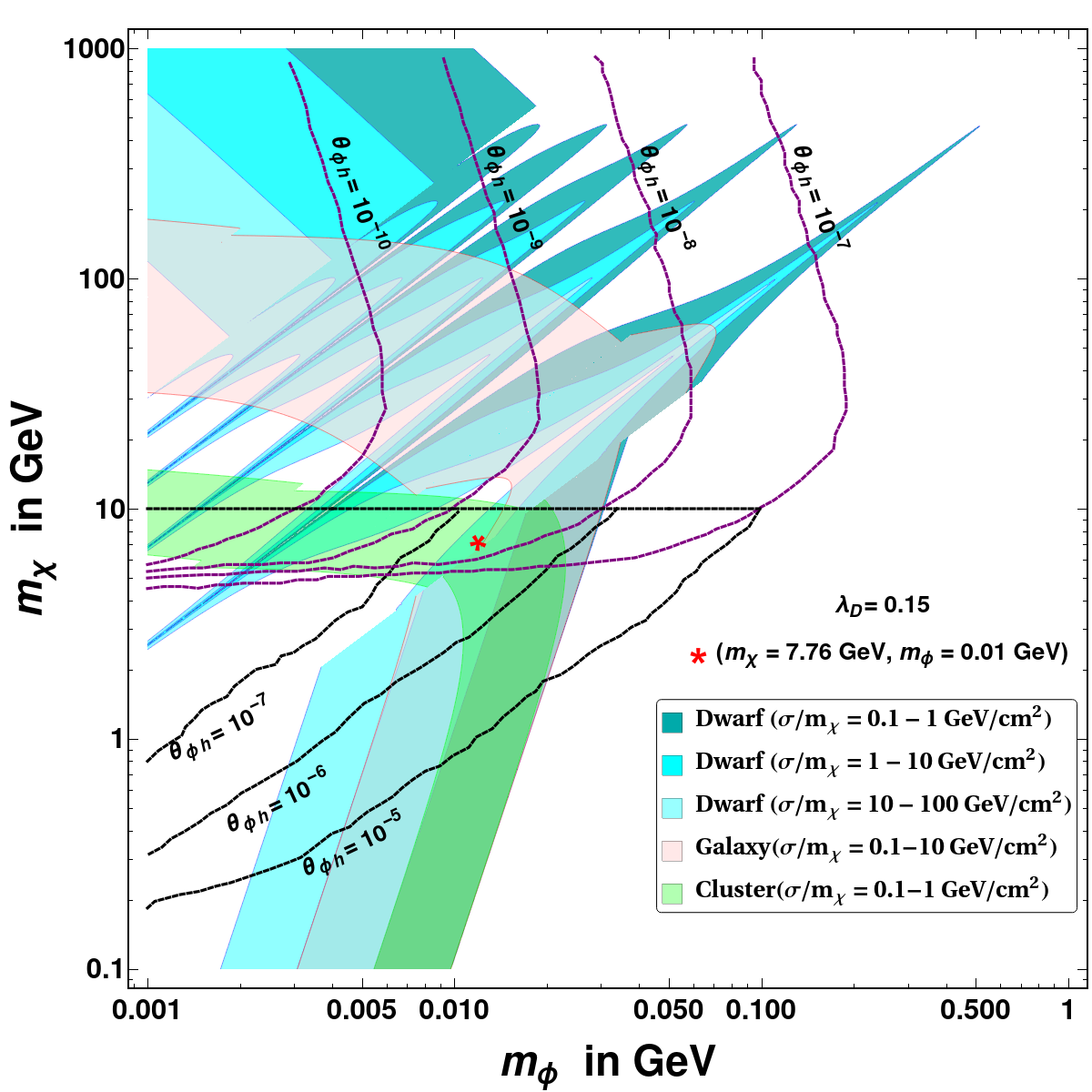}
		\caption{ Self-interaction allowed parameter space constrained by DM direct search in the plane of $(m_{\chi})$ versus $(m_\phi)$.}	
			\label{sidmdd}
	\end{figure}

	The mixing angle $\theta_{\phi h}$ can be derived in terms of the parameters $\lambda_{H \Phi},\langle \Phi \rangle, v, m_\phi, m_h$. Depending on the value of $\lambda_{H \Phi}$, the $\phi - h$ mixing can be very small or large. $\theta_{\phi h}$ gets an upper bound from invisible Higgs decay (since typically $m_\phi < m_h$), while it has a conservative lower bound from the big bang nucleosynthesis (BBN)(since $\tau_\phi <  \tau_{\rm BBN}$ in order to keep BBN predictions intact) \cite{Borah:2021rbx}. 	
	%In Fig. \ref{summary10a}, we have shown the lower bound on $\theta_{\phi h}$ as a function of $m_\phi$ from this lifetime criteria. We see from Fig.\ref{summary10a} that $\theta_{\phi h} < 10^{-11}$ is disfavoured for all $m_\phi \sim 10$ MeV.    
	%
%	\begin{figure}[h!]
%		\includegraphics[width=8cm,height=8cm]{scalar_decay.png}
		%\hfill
		%\includegraphics[width=8cm,height=8cm]{sd_sidm_dd_xenon1t}
	%	\caption{Lifetime of $\phi$ is shown in the plane of $\theta_{\phi h}$ versus $m_\phi$.}% Right: Constraints from DM direct detection in the plane of DM mass $(m_\chi)$ versus mediator mass $(m_\phi)$ for self-interaction.}
%		\label{summary10a}
%	\end{figure}	
	Using Eq.~\eqref{fpn} and \eqref{DD4}, the spin-independent DM-nucleon scattering cross-section in Eq.\,\eqref{DD_cs}, can be re-expressed as:

	\begin{eqnarray}
		\sigma_{\rm SI}^{\phi-h} &=&\frac{{\mu_{r}}^{2} \lambda_D \theta^2_{\phi h} }{\pi A^{2}}  \left[ \frac{1}{m^2_{\phi}}-\frac{1}{m^2_{h}} \right]^{2} \nonumber\\ & \times & \bigg[ Z \left(\frac{m_{p}}{v}\right) \left(f_{Tu}^{p}+f_{Td}^{p}+f_{Ts}^{p}+\frac{2}{9}f_{TG}^{p} \right) \nonumber\\ &+& (A-Z) \left(\frac{m_{n}}{v}\right) \left(f_{Tu}^{n}+f_{Td}^{n}+f_{Ts}^{n}+\frac{2}{9}f_{TG}^{n}\right)  \bigg]^{2}. \nonumber\\
		\label{SI_cross}
	\end{eqnarray}		
	Among the direct search experiments, CRESST-III~\cite{Abdelhameed:2019hmk} provides the most severe constraint on DM mass below 10 GeV, while XENON1T \cite{Aprile:2018dbl} provides the stringent constraints for DM mass above 10 GeV. In Fig.~\ref{sidmdd}, these constraints are shown on the $m_{\chi}-m_\phi$ plane against the self-interaction favoured parameter space. The blue (purple) coloured contours denote exclusion limits from XENON1T (CRESST-III) experiment for specific $\phi-h$ mixing parameter $\theta_{\phi h}$. The region to the left of each contour is excluded for that particular $\theta_{\phi h}$. It is seen from Fig.~\ref{sidmdd} that direct search experiments severely constrain the self-interaction favoured parameter space. In particular, for $m_{\chi}=7.76$ GeV and $m_\phi=0.01$ GeV, $\theta_{\phi h} > 10^{-9}$ has already been ruled out. The red star mark depicts the benchmark point for fully asymmetric DM.
	
\section{Neutrino Mass}\label{neutino_mass}
We explain neutrino mass through the higher dimension operator,
%\begin{equation}
%\mathcal{L}_{Dirac} = -y \frac{\rho}{M_{X}^{2}} \overline{L} H \Phi' \nu_{R}
%\end{equation}

\begin{equation}
\mathcal{L}_{Dirac} = -y_{1} \frac{\rho_{1}}{M_{X_{1}}^{2}} \overline{L} H \Phi' \nu_{R}-y_{2} \frac{\rho_{2}}{M_{X_{2}}^{2}} \overline{L} H \Phi' \nu_{R}
\end{equation} 
 
which is a Dirac-type dimension-5 operator\cite{Gu:2006dc}, where the trilinear coupling $\rho$ has mass dimension. This operator shares the essential features of conventional Majorana-type dimension-5 operator\cite{Weinberg:1979sa,Minkowski:1977sc, Yanagida:1981xy, GellMann:1980vs, Mohapatra:1979ia}. 
\begin{figure}[ht]
				\centering
				\includegraphics[width=5cm,height=4cm]{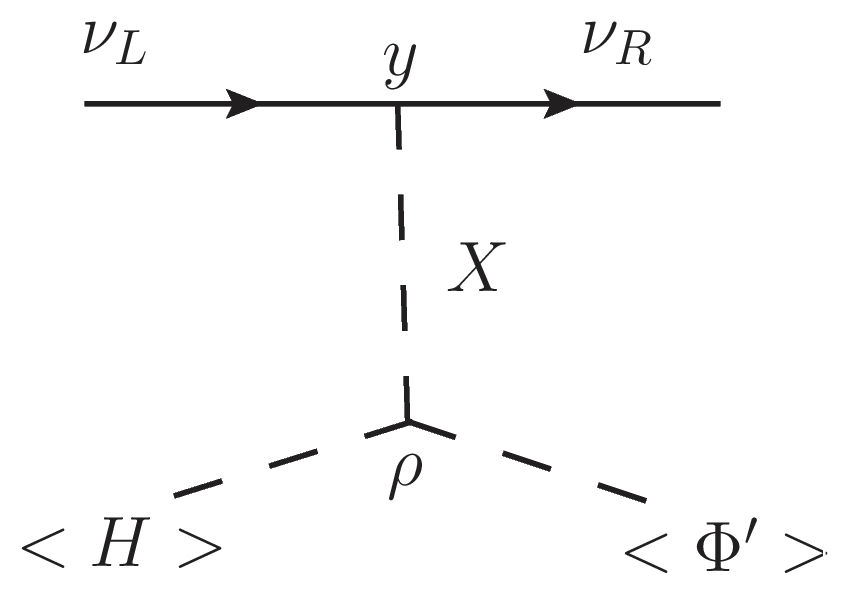}
				 \caption{Dirac neutrino mass generated through dimension-5 operator.}
\end{figure}
We get small Dirac neutrino mass after the SM-Higgs H and $\Phi'$ acquire vev's $\langle H \rangle = v$ and $\langle \Phi' \rangle = w$ respectively.

%\begin{equation}
%M_{\nu} \simeq y \frac{\rho v w}{M_{X}^{2}} .
%\end{equation}

\begin{equation}
M_{\nu} \simeq y_{1} \frac{\rho_{1} v w}{M_{X_{1}}^{2}} + y_{2} \frac{\rho_{2} v w}{M_{X_{2}}^{2}} .
\end{equation}

Again, using the definition $f_{1,2}=\frac{\rho_{1,2}}{M_{X_{1,2}}}$, we can write,

%\begin{equation}
%M_{\nu} \simeq y f^2 v \frac{w}{\rho} .
%\end{equation}

\begin{equation}
M_{\nu} \simeq y_{1} f_{1}^2 v \frac{w}{\rho_{1}} + y_{2} f_{2}^2 v \frac{w}{\rho_{2}} .
\end{equation}

The mass of the heavy scalars $X_{1,2}$ {\it i.e.,} $M_{X_{1,2}}$ is already decided from the requirement of Dirac Leptogenesis to be of $\mathcal{O}(10^{10,11})\,{\rm GeV}$ (see section-\ref{Gen_Asy_DM_transfer}). To explain the neutrino mass of the order of $0.1\,{\rm eV}$, we pick a particular solution for $y_{1,2}$ and $f_{1,2}$, say $y_{1}=2.2 \times 10^{-4}$, $y_{2}=1.1 \times 10^{-4}$ and $f_{1}=7.3\times 10^{-5}$, $f_{2}=1.1$,
%from the available parameter region shown in Fig.\,\ref{Log f vs Log y}, 
which also explains successfully the lepton asymmetry in section-\ref{Gen_Asy_DM_transfer}. The other parameter which appear to achieve neutrino mass of $\mathcal{O}(10^{-10})\,{\rm GeV}$ is $w=10^{2.7}$ GeV. 
%and $\rho$ have to be of the same order as clearly seen from the following calculation.
%
%\begin{equation}. 
%\begin{aligned}
%\frac{w}{\rho} &\simeq \Bigg(\frac{M_\nu}{v}\Bigg)\frac{1}{y f^2}\\
%&= \Bigg(\frac{10^{-10}}{246}\Bigg)\frac{1}{2.2 \times 10^{-4} \times (7.3\times 10^{-5})^2}\\
%               &\simeq \mathcal{O}(1)\\               
%\end{aligned}
%\end{equation}

Thus the parameters $y,\rho$ and $M_X$ are the bridging ligands for leptogenesis and neutrino mass. A typical set of parameters for which the model actually satisfies all the relevant phenomenology simultaneously can be given as: $y_{1}=2.2 \times 10^{-4}$, $y_{2}=1.1 \times 10^{-4}$, $\rho_{1}=7.3\times 10^{5}$ GeV, $\rho_{2}=1.1 \times 10^{11}$ GeV,
$M_{\zeta_{1}} = 10^{10}$ GeV, $M_{\zeta_{2}} = 10^{11}$ GeV, $m_\phi$ = 0.003-0.03 GeV, $\theta_{\phi h} \leq 10^{-9}$, $\omega=10^{2.7}$ GeV, $v=246$ GeV, $u \sim 10^{-2}-10^{-3}$ GeV, $\lambda_H = 0.129, \lambda_\phi' \sim 1,  \mu_1 \sim \mu_2 \sim -10^{-12} {\rm GeV}, \lambda_{H \Phi'} \sim 10^{-3}, \lambda_{H \Phi} \sim 10^{-10}, \lambda_{\Phi \Phi'} \sim 10^{-10}$.  Note that in the above, we use a range of $m_{\phi}$ values (which give rise the correct order of DM self-interaction) for which we get a range of $u$ values.

Corresponding to $\lambda=1.5$ and $\lambda_{\eta H}=1$ (see Eq.\ref{eff_O8}) the mass scale of $\eta$ is $M_{\eta}=1.1 \times 10^{4}$ GeV.

The set of parameters which gives rise to the correct dark matter relic density is given as: $m_\chi$ = 1.3-7.76 GeV, $\lambda_{D}$ = 0.02-0.15. Since we are considering the dominant asymmetric DM, the relic density is solely decided by these two parameters only. 

 %

%%%%%%%%%%%%%%%%%%%%%%%%%%%%%%%%%%%%%%%%%%%%%%%%%%%%%%%%%%%%%%%%%%%%%%%%%%%%%%%%%%%%%%%%%%%%%%%%%%%
\section{Conclusion}\label{conclusion}
%%%%%%%%%%%%%%%%%%%%%%%%%%%%%%%%%%%%%%%%%%%%%%%%%%%%%%%%%%%%%%%%%%%%%%%%%%%%%%%%%%%%%%%%%%%%%%%%%%%
We extended the SM with two super heavy $SU(2)_L$ scalar doublets: $X_{i}$ $(i =1,2)$, three right-handed neutrinos: 
$\nu_{R_{i}} (i=1, 2, 3)$, two singlet scalars: $\Phi$, $\Phi'$ and a singlet Dirac fermion $\chi$ which represents the candidate of 
a self-interacting DM. We assumed a mass hierarchy among the heavy scalar doublets $X_i$, $i=1,2$. As a result the CP-violating 
out-of-equilibrium decay of the lightest heavy scalar generated a neutrino asymmetry in the visible sector. A part of the neutrino asymmetry 
is then transferred to the dark sector by a dimension-8 operator: $\mathcal{O}_{8}$, while a part of the remaining neutrino asymmetry gets converted to 
the baryon asymmetry by the Electroweak sphaleron processes. This asymmetry transfer mechanism establishes a proportionality between the relic 
densities of dark matter and baryonic matter. The ratio between these two relic densities fixes the mass of the DM to be 7.76 GeV, if DM relic 
is fully asymmetric. However depending on the fractional contribution from symmetric and the asymmetric components, the DM mass can vary in the 
range $0.07-7.76$ GeV. The light scalar mediator $\phi$ is introduced, not only facilitate velocity-dependent DM self-interaction to alleviate 
small-scale issues of $\Lambda{\rm CDM}$, but also deplete the symmetric component of DM via efficient annihilation process: $\overline{\chi} \chi 
\to \phi \phi$. The requirement of sufficient self-interaction also rules out a region of available parameter space in terms of DM mass, further 
restricting it to $1.3-7.76$ GeV. To realise sufficient DM self-interaction with this mass range of DM, the light scalar mass must be 
in the range $0.003-0.03$ GeV. The self-interaction allowed parameter space has also been confronted with bounds from early Universe physics 
like BBN and present Universe physics like DM direct search and allowed parameter space from all phenomenological constraints has been specified. 
We have also explained the small Dirac neutrino mass at tree-level through a dimension-five operator, where the mass of the heavy scalar $X$ suppresses 
the weak scale. $M_{X_{1}}$ is decided from the requirement of Dirac Leptogenesis to be of $\mathcal{O}(10^{10})\,{\rm GeV}$ (see \ref{Gen_Asy_DM_transfer}). 
To generate the desired lepton asymmetry, the branching ratio $B_L(X \to L\nu_R) \sim 0.9$, which requires that $f_{1} =\frac{\rho_{1}}{M_{X_{1}}}$ and $y_{1}$ differ 
roughly by an order of magnitude. A typical solution $y_{1}=2.2 \times 10^{-4}$, $y_{2}=1.1 \times 10^{-4}$ and $f_{1}=7.3\times 10^{-5}$, $f_{2}=1.1$ simultaneously satisfies both lepton asymmetry 
and neutrino mass. Therefore, the model at hand successfully explains the baryon asymmetry of the Universe along with neutrino mass and DM relic while 
providing a solution for small scale $\Lambda{\rm CDM}$ anomalies via DM self-interaction.

%%%%%%%%%%%%%%%%%%%%%%%%%%%%%%%%%%%%%%%%%%%%%%%%%%%%%%%%%%%%%%%%%%%%%%%%%%%%%%%%%%%%%%%%%%%%%%%%%%%
\section*{Acknowledgement}
\label{Acknowlegement}
MD acknowledges SERB CORE Grant CRG/2018/004971. NS acknowledges the support from Department of Atomic Energy (DAE)-Board of Research in Nuclear Sciences (BRNS), Government of 
India (Ref. Number: 58/14/15/2021-BRNS/37220).

\appendix

%\section*{Appendix}
\section{Asymmetry transfer from visible to dark sector}\label{appendix_asy}
\label{chemical}

The asymmetry in the equilibrium number densities of particle $n_{i}$ over antiparticle $\overline{n}_{i}$ can be written as,
\begin{equation}
n_{i}-\overline{n}_{i} = \frac{g_{i}}{2 \pi^{2}} \int_{0}^{\infty}dq ~ q^{2} \left[ \frac{1}{e^{\frac{E_{i}(q)-\mu_{i}}{T}}\pm 1}-\frac{1}{e^{\frac{E_{i}(q)+\mu_{i}}{T}}\pm 1} \right]\,,
\end{equation}
where the $g_{i}$ is the internal degrees of freedom of the particle species $i$. In the above equation $E_{i}$ and $q_{i}$ represent 
the energy and momentum of the particle species $i$. In the approximation of a weakly interacting plasma, where $\beta \mu_{i} \ll 1$, $\beta \equiv 1/T$ (detailed discussions are given in~\cite{Feng:2012jn,Kolb:1990vq}) we get, 
\begin{eqnarray}
\label{aa2}
n_{i}-\overline{n}_{i} & \sim & \frac{g_{i}T^{3}}{6} \times [ 2\beta\mu_{i}+\mathcal{O}\left((\beta \mu_{i})^{3}\right)~~~~ {\rm for \,\,bosons} \nonumber\\
& \sim & \frac{g_{i}T^{3}}{6} \times [ \beta \mu_{i}+\mathcal{O}\left((\beta \mu_{i})^{3}\right) ~~~~~{\rm for \,\,fermions}. \nonumber\\
\end{eqnarray}
In this model, the asymmetry transfer operator is given by ${\cal O}_8= \frac{1}{M_{asy}^{4}} \overline{\chi}^{2}(LH)^{2}$ so the decoupling temperature of operator ${\cal O}_8$ is depends on the value of $M_{\rm asy}$. Since in this model the $B-L$ asymmetry is generated in the standard model sector which is required to be shared to the dark sector via ${\cal O}_8$ operator, we assume the decoupling temperature $T_D$ to be $T_{t} > T_{D} > T_{W}$, where $T_{t}$ is the temperature of thermal bath when the top quark decouples and $T_{W}$ is the temperature when the $W$ boson decouples from the thermal plasma. In this case the effective Lagrangian for Yukawa coupling is given by:
\begin{equation}
\mathcal{L}_{Yukawa} = g_{e_{i}}^{k}\bar{e}_{iL}h^{k}e_{iR}+g_{u_{i}}^{k} \bar{u}_{iL}h^{k}u_{iR}+g_{d_{i}}^{k}\bar{d}_{iL}h^{k}d_{iR} +h.c
\label{b1}
\end{equation}
where $k=1,2,3$ for three scalar mass eigenstates $h_{1}$, $h_{2}$ and $h_{3}$. Where $h_{1}$ is identified as standard model Higgs boson with mass $125$ GeV. All these scalar fields are real so the above Lagrangian gives the following chemical equilibrium condition,
\begin{equation}
0=\mu_{h^{k}}=\mu_{u_{L}}-\mu_{u_{R}}=\mu_{d_{L}}-\mu_{d_{R}}=\mu_{e_{L}}-\mu_{e_{R}}.
\label{b2}
\end{equation}
After electroweak symmetry breaking the charged current interaction part of the SM Lagrangian is given by,
\begin{equation}
\mathcal{L}_{int}^{(W)}=gW_{\mu}^{+}\bar{u}_{L}\gamma^{\mu} d_{L} + gW_{\mu}^{+}e_{L}\gamma^{\mu}\bar{\nu}_{eL}\,.
\label{b3}
\end{equation}
The charged current interactions remain in thermal equilibrium until W-boson decouples from thermal bath. Which gives the following chemical potential equations,
\begin{equation}
\mu_{W}=\mu_{u_{L}}-\mu_{d_{L}},
\label{b4}
\end{equation}
and
\begin{equation}
\mu_{W}=\mu_{\nu}-\mu_{e_{L}}.
\label{b5}
\end{equation}
The electroweak sphalerons remain in thermal equilibrium until a temperature $T_{\rm sph} \gtrsim  T_{W}$ leads to the following constraint,
\begin{equation}
\mu_{u_{L}}+2\mu_{d_{L}}+\mu_{\nu}=0.
\label{b6}
\end{equation}

At a temperature below electroweak phase transition, the electric charge neutrality of the Universe holds.  However, 
at the epoch: $T_{t} > T_{D} > T_{W}$, the top  quark is already decoupled from the thermal plasma and hence 
does not take part in the charge neutrality condition. Therefore, we get 
\begin{equation}
Q=4(\mu_{u_{L}}+\mu_{u_{R}})+6\mu_{W}-3(\mu_{d_{L}}+\mu_{d_{R}}+\mu_{e_{L}}+\mu_{e_{R}})=0.
\label{c1}
\end{equation}

Using Eq~\ref{aa2}, the baryon number density $n_{B}$ can be given as,
\begin{equation}
n_{B} = \sum_{i} \mu_i \big(C_{i} Q_{i}^{B}\frac{g_i T^3 \beta}{6}\big) 
\end{equation}
where, $i$ runs over two generations of up quarks and three generations of down quarks, $C_{i}$ and $Q_{i}^B$ count respectively the color and the baryon number of $i^{th}$ quark. Note that top quark is decoupled since it is heavy. Here, $g_i = 2$ is the internal degrees of freedom of each quark.
Similarly, following Eq~\ref{aa2}, the lepton number density $n_{L}$ can be given as,
\begin{equation}
n_{L} = \sum_{i} \mu_i \big(\frac{g_i T^3 \beta}{6}\big) 
\end{equation} 
where, $i$ runs over three generations of charged and neutral leptons.
Now using the Eqs.\,\ref{b2}-\ref{c1}, we can write the net baryon and lepton number density $n_{B}$ and $n_{L}$ as,
\begin{equation}
n_{B}=-\frac{90}{29}\mu_{\nu}
\label{c2}
\end{equation}
and 
\begin{equation}
 n_{L}=\frac{201}{29}\mu_{\nu}\,,
\label{c3}
\end{equation}
where we have dropped the common factor $g T^3\beta/6$ as we are interested in ratio of 
densities, rather than their individual values. The net $B-L$ asymmetry in the visible sector is thus given by,
\begin{equation}
(n_{B-L})_{\rm vis}=-\frac{291}{29}\mu_{\nu}\,.
\label{c4}
\end{equation}
After sphaleron processes decouple at $T_{\rm sph}$, the baryon and lepton number densities would be conserved separately. 
As a result Eqs.\,\ref{b2} -~\ref{c4} would remain valid at 
$T_{\rm sph} > M_{W}$. Once the sphaleron processes decouple, the ratio of $n_{B}/n_{B-L}$ would be frozen. As 
a result from Eqs.\,\ref{c2}  and \ref{c4}, it can be written as,
\begin{equation}
\frac{n_{B_{\rm final}}}{(n_{B-L})_{\rm vis}}=\frac{n_{B}}{(n_{B-L})_{\rm vis}}=\frac{30}{97}=0.31
\label{c5}
\end{equation}
\begin{equation}
n_{B_{\rm final}}=0.31(n_{B-L})_{\rm vis}.
\label{c6}
\end{equation}

%As the asymmetry transfer operator ${\cal O}_8$ is active down to W-boson decoupling temperature, a part of the $B-L$ 
%asymmetry gets transferred from visible sector to the dark sector. Therefore, $(n_{B-L})_{\rm vis}$ is no longer equal 
%to the $(n_{B-L})_{\rm total}$. 

The operator ${\cal O}_8$ is in equilibrium until $T_{D} > T_{\rm sph}$ and equilibration of ${\cal O}_8$ gives the following constraint, 
\begin{equation}
\mu_{\chi} = \mu_{\nu} \,.
\label{c7}
\end{equation}
As a result the number density of dark matter $\chi$ (it is basically $B-L$ asymmetry in the dark sector) is given by, 
is given by,

\begin{equation}
\begin{aligned}
n_{\chi}&= 2\mu_{\chi} \frac{g T^3 \beta}{6} \\ &= \frac{58}{291}(n_{B-L})_{\rm vis} \equiv (n_{B-L})_{\rm dark}
\end{aligned}
\label{c8}
\end{equation}

\section{Diagonalization of the Scalar Mass Matrix}
	\label{appendix_diag}
	
	For simplicity, let us denote the scalar mass matrix given in Eq.\,\ref{matrix_scalar} as:
\begin{equation}
\begin{pmatrix}
A & B & C&\\
B & D & E&\\
C & E & F&\\
\end{pmatrix}
\end{equation} 
This mass matrix can be approximately block-diagonalized by the matrix:
\begin{equation}
U= \begin{pmatrix}  
 \cos \alpha    &   0  &    \sin\alpha   & \\
 0     &  \cos\alpha  & \sin\alpha &  \\ 
 -\sin \alpha   &    -\sin \alpha &    \cos\alpha   & \\
   \end{pmatrix}
\label{u_matrix}
\end{equation}
which is unitary upto order $\sin\alpha$. The block diagonalized matrix is of the form:
\begin{equation}
\begin{pmatrix}
A' & B' & 0&\\
B' & D' & 0&\\
0 & 0 & F'&\\
\end{pmatrix}
\label{b_matrix}
\end{equation}
where the elements of the above matrix are as follows:
\begin{equation}
\begin{aligned}
A'&=A \cos^2\alpha + F -\sin^2\alpha - C \sin2\alpha
\\
B'&=B \cos^2\alpha + F \sin^2\alpha - \frac{1}{2}(C+E)\,\sin2\alpha
\\
D'&=D \cos^2\alpha + F \sin^2\alpha - E \sin2\alpha
\\
F'&=F \cos^2\alpha + (A+2B+D)\,\sin^2\alpha + (C+E)\,\sin2\alpha
\end{aligned}
\end{equation}
The above diagonalization is obtained for the values of the angle $\alpha$ given by(ignoring $\mathcal{O}(\sin^2\alpha)$ terms):
\begin{equation}
\tan\alpha =  \frac{C}{F-A-B} ~~~~{\rm (or)}~~~~ \frac{E}{F-B-D} \,\,.
\label{tanalpha}
\end{equation}
We assume the mass of $\phi$ to be extremely small compared to that of both $\phi'$ and $h$ and it has very small mixing with both of them. We consider the induced vev u of the field $\Phi$ to be very small compared to that of SM Higgs. In the limit of zero $\phi'-\phi$ as well as $h-\phi$ mixing, which is indeed guaranteed by the extremely tiny mixing parameter $\tan\alpha$ given by Eq.\,\ref{tanalpha}, $\phi$ decouples from both $\phi'$ and h. The smallness of $\tan \alpha$ can be understood with the help of parameter values provided in Sec.-\ref{neutino_mass} {\it viz.} $\omega=10^{2.7}$ GeV, $v=246$ GeV, $u \sim 10^{-2}$ GeV, $\lambda_\phi' \sim 1, \lambda_H = 0.129$, $\mu_2 \sim -10^{-12}$, $\lambda_\phi' \sim 1, \lambda_{H \Phi'} \sim 10^{-3}, \lambda_{H \Phi} \sim 10^{-10}, m_\phi \sim 10^{-2}$ GeV, which gives $\tan \alpha \sim \mathcal{O} (10^{-12})$.
%$\lambda_H = 0.129, \lambda_\phi' \sim 1,  \mu_1 \sim \mu_2 \sim -10^{-12}, \lambda_{H \Phi'} \sim 10^{-3}, \lambda_{H \Phi} \sim 10^{-10}, \lambda_{\Phi \Phi'} \sim 10^{-10}$}
%Only the (1,2) block of the mass matrix given in Eq.\,\ref{scalar_mass_matrix} is sufficient to study the low energy phenomenology. In order to show small (1,3) and (2,3) mixings, we set $m_\Phi= 0  \,{\rm GeV}$ and consider the parameter set as $v=246 \,{\rm GeV}, w=10^{5} \,{\rm GeV}, u=1 \,{\rm GeV}, \mu_1=1 \,{\rm GeV}, \mu_2=10^{-2} \,{\rm GeV}, \lambda_H= 0.13, \lambda_\Phi=1, \lambda_\Phi'=1, \lambda_{H\Phi}=0.1, \lambda_{H\Phi'}=10^{-5} \,\,{\rm and}\,\, \lambda_{\Phi \Phi'}=10^{-7}$. The (1,3) mixing is specified by $\tan\alpha = \frac{C}{F-A-B}\approx 1.7\times 10^{-8}$, while the (2,3) mixing is specified by  $\tan\alpha = \frac{E}{F-B-D}\approx 6.15\times 10^{-9}$. Due to such extremely small mixing, the masses of $\phi$ and h are not affected in spite of $\phi'$ having very high mass. 
The exact diagonalization is obtained by giving a consecutive (1,2) rotation to the squared mass matrix given in Eq.\,\ref{b_matrix} by the following Euler rotation matrix:
\begin{equation}
O=\begin{pmatrix}
\cos \beta	&		-\sin \beta			& 	 	  0		  & \\
\sin \beta	&	\cos \beta 	&	0   & \\
0	&	0 	&	1   & \\
\end{pmatrix}
\label{o_matrix}
\end{equation}
where the $\phi' - h$ mixing is given by,
\begin{equation}
\tan 2\beta = \frac{2B'}{D'-A'}
\end{equation}
After complete diagonalization, we are left with an extremely light scalar $h_3 \approx \phi$ that mediates DM self-scattering and two other scalars with masses given by,
\begin{equation}
\begin{aligned}
m^2_{h_1}&= D' \cos^2\beta + A' \sin^2\beta + B\sin2\beta 
\\
m^2_{h_2}&= A' \cos^2\beta + D' \sin^2\beta - B\sin2\beta  ~.
\end{aligned}
\end{equation}
%The physical mass eigenstates are related to the unphysical states through product of the two unitary rotation matrices U and O given by Eq.\,\ref{u_matrix} and Eq.\,\ref{o_matrix} respectively. However, in the small $\alpha$ limit which is indeed the case here, the Orthogonal rotation matrix given by Eq.\,\ref{o_matrix} is sufficient and we can safely write,
%\begin{equation}
%\begin{pmatrix}
%h_{1}  & \\
%h_{2}      & \\
%h_{3}   & \\
%\end{pmatrix}=\begin{pmatrix}
%\cos\beta&	-\sin\beta& 	 0  \\
%\sin\beta	&	\cos\beta	&	0   & \\
%0	&	0 	&	1  & \\
%\end{pmatrix}
%\begin{pmatrix}
%\phi  & \\
% h     & \\
%\phi'   & \\
%\end{pmatrix}.
%\end{equation}
The mass eigenstate $h_1$ can be identified as the SM Higgs with mass $M_{h_{1}}=125.18 \,{\rm GeV}$, while $h_2$ as the second scalar that plays a role in generating the tiny neutrino mass.
\section{DM Self-interaction Cross-sections at Low Energy}
	\label{appendix2}
	In the Born Limit ($\lambda^2_D m_{\chi}/(4\pi m_\phi) << 1$),
	\begin{equation}
		\sigma^{Born}_T=\frac{\lambda^4_D}{2\pi m^2_{\chi} v^4}\Bigg(ln(1+\frac{ m^2_{\chi} v^2}{m^2_\phi})-\frac{m^2_{\chi}v^2}{m^2_\phi+ m^2_{\chi}v^2}\Bigg)
	\end{equation} 	
	Outside the Born regime ($\lambda^2_D m_{\chi}/(4\pi m_\phi) \geq 1 $), there are two distinct regions {\it viz}, the classical regime and the resonance regime. In the classical regime ($\lambda^2_D m_{\chi}/(4\pi m_\phi \geq 1, m_{\chi} v/m_{\phi} \geq 1$), the solutions for an attractive potential is given by\cite{Tulin:2013teo,Tulin:2012wi,Khrapak:2003kjw}:	
	\vspace*{0.5cm}	
	\begin{equation}
		\sigma^{classical}_T =\left\{
		\begin{array}{l}			
			\frac{4\pi}{m^2_\phi}\beta^2 ln(1+\beta^{-1}) ~~~~~~~~~~~\beta \leqslant 10^{-1}\\
			\frac{8\pi}{m^2_\phi}\beta^2/(1+1.5\beta^{1.65}) ~~~~~~~~10^{-1}\leq \beta \leqslant 10^{3}\\
			\frac{\pi}{m^2_\phi}( ln\beta + 1 -\frac{1}{2}ln^{-1}\beta) ~~~~~\beta \geq 10^{3}\\
		\end{array}
		\right.
	\end{equation}  \\	
	where $\beta = 2 \lambda^2_D m_{\chi}/(4\pi m_\phi) v^2$	
	%Both the Born and the classical regime do not provide us the mild velocity dependence in the cross-section required to explain the small scale issues. However one interesting 	
	In the resonant regime ($\lambda^2_D m_{\chi}/(4\pi m_\phi) \geq 1, m_{\chi} v/m_{\phi}  \leq 1$), the quantum mechanical resonances and anti-resonance in $\sigma_T$ appear due to (quasi-)bound states formation in the attractive potential. In the resonant regime, an analytical formula for $\sigma_T$ is not available and one needs to solve the non-relativistic Schrodinger equation by partial wave analysis. Instead, here we use the non-perturbative results obtained by approximating the Yukawa potential to be a Hulthen potential $\Big( V(r) = \pm \frac{\lambda^2_D}{4\pi}\frac{ \delta e^{-\delta r}}{1-e^{-\delta r}}\Big)$, which is given by~\cite{Tulin:2013teo}:	
	\begin{equation}
		\sigma^{Hulthen}_T = \frac{16 \pi \sin^2\delta_0}{m^2_{\chi} v^2}
	\end{equation}
	where $l=0$ phase shift $\delta_0$ is given in terms of the $\Gamma$ functions by :
	\begin{eqnarray}
		\delta_0 &=arg \Bigg(i\Gamma \Big(\frac{i m_\chi v}{k~ m_\phi}\Big)\bigg/{\Gamma (\lambda_+)\Gamma (\lambda_-)}\Bigg)\nonumber\\
		\lambda_{\pm} &=
		\begin{array}{l}
			1+ \frac{i m_\chi v}{2 ~k ~m_\phi}  \pm \sqrt{\frac{\alpha_D m_\chi}{k m_\phi} - \frac{ m^2_\chi v^2}{4 k^2 m^2_\phi}}\\
		\end{array}
	\end{eqnarray}   
	and $k \approx 1.6$ is a dimensionless number.

	%\newpage
	
%\end{document}
%\bibliographystyle{JHEP}
%\bibliography{ref}
%\providecommand{\href}[2]{#2}\begingroup\raggedright

%%%%%%%%%%%%%%%%%%%%%%%%%%%%%%%%%%%%%%%%%%%%%%%%%%%%%%%%%%%%%%%%%%%%%%%%%%%%%%%
%\bibliography{ADM,ref22,refrhn,refn,ref,ref_old,ref1,ref2,ref3,refl,ref11,ref12,re13,ref14}
%\end{thebibliography}
\end{document}